\documentclass[a4paper,fleqn]{cas-sc}

\usepackage[numbers,sort&compress]{natbib}
\usepackage{lineno,hyperref}
\usepackage{subfigure}
\usepackage{booktabs}
\usepackage{amsmath}
\usepackage{amssymb}
\usepackage{numcompress}
\usepackage{multirow}
\usepackage{multicol}
\usepackage{bbding}
\usepackage{tabularx}
\usepackage{makecell}

\def\tsc#1{\csdef{#1}{\textsc{\lowercase{#1}}\xspace}}
\tsc{WGM}
\tsc{QE}

\begin{document}

\let\WriteBookmarks\relax
\def\floatpagepagefraction{1}
\def\textpagefraction{.001}

\shorttitle{DIICAN: Dual Time-scale State-Coupled Co-estimation of SOC, SOH and RUL for Lithium-Ion Batteries}    

\title [mode = title]{Dual time-scale state-coupled co-estimation of SOC, SOH and RUL for lithium-ion batteries via Deep Inter and Intra-Cycle Attention Network}  

\author[1]{Ningbo Cai}
\author[1]{Yuwen Qin}
\author[1]{Xin Chen}
\cormark[1]
\author[2]{Kai Wu}

\affiliation[1]{organization={
Center of Nanomaterials for Renewable Energy, \\ State Key Laboratory of Electrical Insulation and Power Equipment, \\School of Electrical Engineering, \\ Xi'an Jiaotong University},
                postcode={Xi'an 710049}, 
                country={China}}

\affiliation[2]{organization={State Key Laboratory of Electrical Insulation and Power Equipment, \\School of Electrical Engineering, \\ Xi'an Jiaotong University},
                postcode={Xi'an 710049}, 
                country={China}}
\cortext[1]{Corresponding author, Xin Chen, xin.chen.nj@xjtu.edu.cn}

\begin{abstract}
 Accurate co-estimations of battery states, such as state-of-charge (SOC), state-of-health (SOH,) and remaining useful life (RUL), are crucial to the battery management systems to assure safe and reliable management. Although the external properties of the battery charge with the aging degree, batteries' degradation mechanism shares similar evolving patterns. Since batteries are complicated chemical systems, these states are highly coupled with intricate electrochemical processes. A state-coupled co-estimation method named Deep Inter and Intra-Cycle Attention Network (DIICAN) is proposed in this paper to estimate SOC, SOH, and RUL, which organizes battery measurement data into the intra-cycle and inter-cycle time scales. And to extract degradation-related features automatically and adapt to practical working conditions, the convolutional neural network is applied. The state degradation attention unit is utilized to extract the battery state evolution pattern and evaluate the battery degradation degree. To account for the influence of battery aging on the SOC estimation, the battery degradation-related state is incorporated in the SOC estimation for capacity calibration. The DIICAN method is validated on the Oxford battery dataset. The experimental results show that the proposed method can achieve SOH and RUL co-estimation with high accuracy and effectively improve SOC estimation accuracy for the whole lifespan.
\end{abstract}

\begin{highlights}
\item The Deep Inter and Intra-Cycle Attention Network (DIICAN) method is proposed for the co-estimation of SOC, SOH, and RUL.
\item Convolutional neural networks are applied to automatically extract battery degradation-related embedding features from the raw BMS streaming data.
\item The state degradation attention unit is combined with recurrent neural networks to extract the battery state evolving pattern for the SOH and RUL estimation over the whole lifespan.
\item SOH and SOC are coupled to account for the influence of battery aging on the SOC estimation. The SOC estimation accuracy is improved significantly over the battery lifespan.
\end{highlights}

\begin{keywords}
  Lithium-ion battery\sep
  Remaining useful life (RUL)\sep 
  State of health (SOH)\sep
  State of charge (SOC)\sep 
  Attention mechanism\sep
\end{keywords}

\maketitle

\section{Introduction}                                    
Due to the increasing shortage of resources, environmental pollution\cite[]{bonsu2020towards} \cite[]{mayer2021efficiency}, the past decade has witnessed that the generation of electricity is rapidly increasing from non-predictable and variable renewable energy sources\cite{9386520}. Lithium-ion batteries with superior power and energy density, durability, and environmental protection have been widely applied in energy storage, and power systems such as water power, thermal power, wind power, and solar power stations, and so on. A high-efficiency battery management system (BMS) is usually deployed \cite[]{duan2020online} to facilitate a safe and wide range of battery operations. Accurate battery state-of-charge (SOC), state-of-health (SOH), and remaining useful life (RUL) estimation are key modules within BMS for ensuring the reliability, durability, and performance of batteries. 

Generally speaking, the SOC estimation methods include basic, data-driven, and model-based methods. Basic methods include the looking-up table\cite{tian2021electrode} and Ampere-hour integral methods\cite[]{wang2021improved}. Basic methods have been widely used in practical engineering due to the advantages of simple calculation and easy implementation, but are greatly affected by the working environment \cite[]{xiong2020lithium}, rest time\cite[]{zhang2022novel} and accuracy of the initial value. The model-based method is to study the relationship between the internal mechanism and the external state, establish the model's discrete expression, and then estimate the battery SOC recursively. The method generally owns the merits of real-time and closed-loop feedback. Commonly used models can be roughly summarized into three types: electrochemical models (EM) \cite[]{gao2021co}, equivalent circuit models (ECM) \cite[]{naseri2021enhanced}, and electrochemical impedance models (EIM) \cite[]{wang2020fractional}. Although significant progress has been made in model-based methods, they rely on accurate prior knowledge of internal mechanisms, which is usually unavailable owing to the battery's complex physical and chemical processes, as well as the noise and the diversity of the environment. Data-driven models depend only on historical data and do not need complicated equivalent or mathematical models. \citet[]{hong2020online} proposed an LSTM-based method for multi-forward-step SOC prediction for battery systems in real-world electric vehicles. \citet[]{terala2022state} used combined stacked bi-directional LSTM and encoder-decoder bi-directional long short-term memory architecture to improve on the existing methods of SOC estimation. In \cite[]{chen2022improved}, a GRU model was proposed for accurate SOC estimation under dynamic driving conditions to solve the problems of time long-term dependencies and gradient disappearance or explosion.

Battery aging, usually in the form of capacity fade and resistance growth, is one of the most challenging issues for system safety. Typically, SOH refers to the current health condition of a LIB compared to its initial degradation state. By contrast, RUL is defined as a remaining lifespan from the current cycle to the end of life (EOL) based on its current degradation state. Some model-based approaches \cite[]{bian2020open,sadabadi2021prediction,bi2020online,couto2019state} are mainly to analyze the physical and chemical principles of internal degradation mechanism and establish mathematical models to characterize the process of capacity degradation for SOH and RUL prediction. Numerous methods for extracting health indicators (HIs) have been explored in the recent literature. Indirect HIs extraction methods typically find hidden variation laws and statistical information during the battery operating process. \citet[]{sun2022data} combined incremental capacity analysis (ICA) and bidirectional long short-term memory (Bi-LSTM) neural networks based on health characteristic parameters to predict the SOH of lithium-ion batteries. \citet[]{sun2021battery} used battery terminal voltage during the later stage of the charging process as the input of the sparse auto-encoder and abstracted compressive feature of battery voltage to obtain battery SOH, achieving a good accuracy with adaptability to the capacity fading diversity and voltage differences among different battery cells. \citet{hong2020towards} proposed a dilated CNN-based neural network architecture for predicting the remaining useful life of lithium-ion batteries, which boosted the remaining useful life prediction. \citet[]{deng2022battery} used features extracted from discharge capacity curves to achieve degradation pattern recognition and transfer learning, which can effectively improve SOH estimation accuracy. These relevant research results are flexible and accurate and clearly reflect the advantages of the data-driven method.
\begin{figure}[htbp]
  \centering
  \includegraphics[width=0.85\textwidth]{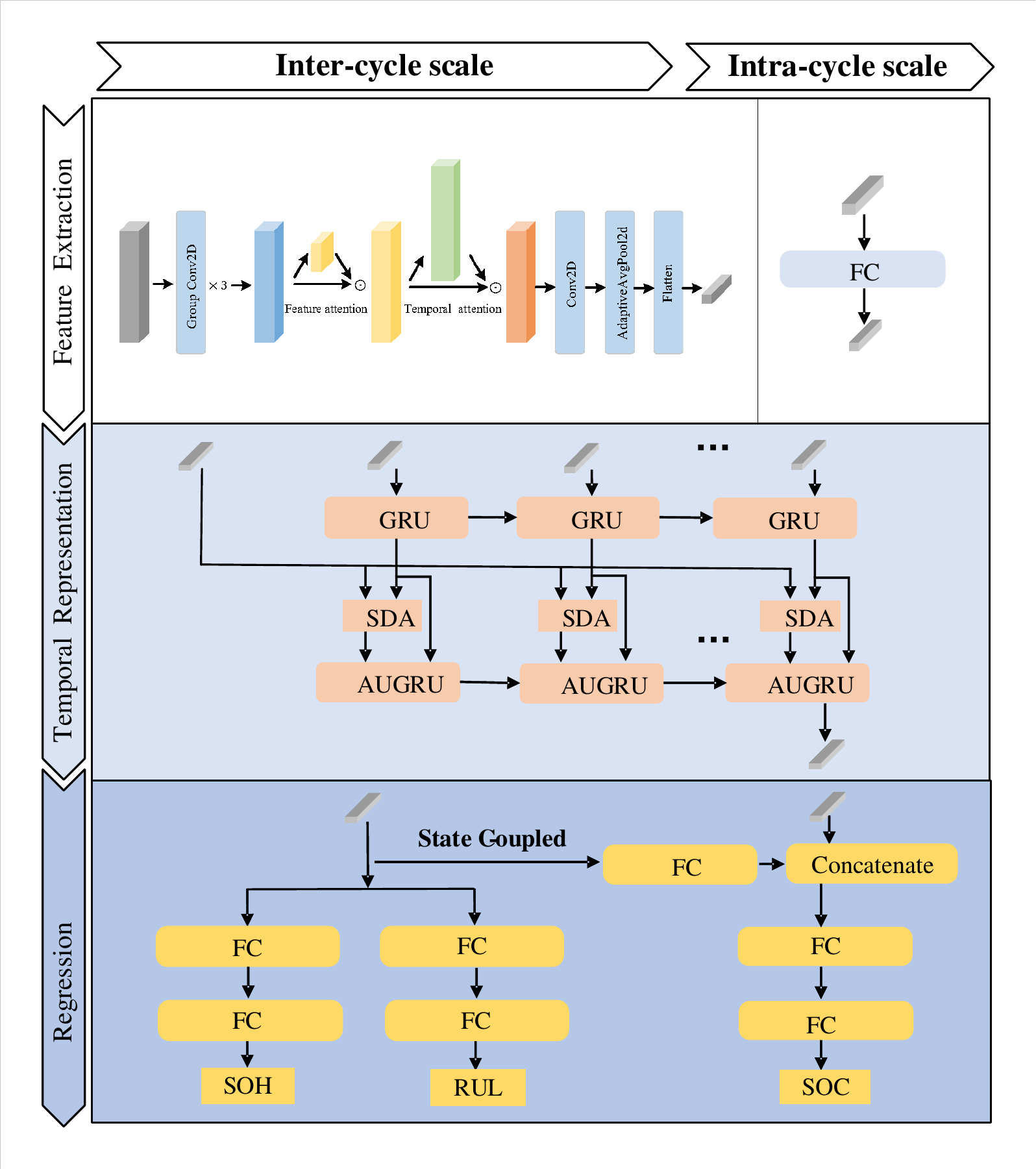}
  \caption{The overall architecture of the DIICAN co-estimation method.}
  \label{fig:framework}
\end{figure}

Monitoring SOC, SOH, and RUL over time is a challenging goal since SOC and SOH are highly coupled with intricate electrochemical processes. With the aging of the battery, the capacity of the battery will gradually decrease, and the characteristics of external measurable parameters may change, which will pose a great challenge to the estimation of SOC. \citet[]{zou2016multi} proposed an effective multi-time-scale estimation algorithm for a class of nonlinear systems with coupled fast and slow dynamics using the developed reduced-order battery models. \citet[]{hu2018co} proposed a SOC and SOH co-estimation scheme that is capable of predicting the voltage response in the presence of initial deviation, noise, and disturbance against battery degradation. \citet[]{che2020soc} established an improved dynamic recurrent neural network (DRNN) with the ability of dynamic mapping to improve the estimation accuracy of the SOC and SOH under different conditions. \citet[]{song2020hybrid} proposed a joint lithium-ion battery state estimation approach with high accuracy and robustness that takes advantage of the least-square-support-vector-machine and unscented-particle-filter.

Some feature extraction methods often require additional manpower consumption, such as incremental capacity analysis (ICA)\cite[]{sun2022data}, differential voltage analysis (DVA)\cite[]{wang2016board} and differential thermal voltammetry (DTV) \cite[]{tian2020state}. Although these HFs have been proven to be highly related to the battery aging process, their adequacy and availability under different working conditions should be considered. In addition, capacity degradation will significantly decrease the accuracy of state estimation, and the accurate state of charge relies on the correction of the maximum available capacity of the battery. The unified data-driven co-estimation method for SOC, SOH, and RUL of battery is crucial work for the modern BMS. To solve the above problems, a dual time-scale state-coupled co-estimation method is adopted in this research, named Deep Inter and Intra-Cycle Attention Network (DIICAN). The key contributions of the present work are summarized as follows:
\begin{itemize}
  \item  The unified state-coupled co-estimation method, DIICAN, is proposed to estimate SOC, SOH, and RUL in battery life cycles according to the inter-cycle and intra-cycle features. The inter-cycle features are used for the SOH-RUL estimation, while the intra-cycle features are for the estimation of SOC. 
  \item Convolutional neural networks are utilized to map raw battery measurements directly to battery degradation-related embedding features automatically to reduce the error caused by manual feature extraction. 
  \item The state degradation attention unit accurately reveals the battery state evolving patterns to represent the battery degradation and achieve the SOH and RUL co-estimation over the whole lifespan. 
  \item The influence of battery degradation on SOC estimation is considered. Battery degradation-related state in the SOH estimation is used for the capacity calibration in the SOC estimation. The accuracy of SOC estimation is improved significantly over the battery lifespan.
\end{itemize} 

The paper is organized as follows. In Sec.~\ref{sec:method}, the proposed DIICAN method is presented. The training process of the DIICAN method is discussed in Sec.~\ref{sec:training}. The experimental results and the performance analysis are given in Sec.~\ref{sec:results}. Finally, Sec.~\ref{sec:conclusion} gives the concluding remarks.

\section{Dual time-scale state-coupled co-estimation: Deep Inter and Intra-Cycle Attention Network}\label{sec:method}
The BMS streaming data has two temporal structures, inter-cycle and intra-cycle timescales. The inter-cycle sequence contains all the information for SOH and RUL while the intra-cycle time series data is used for the estimation of SOC within a cycle. The Deep Inter and Intra-Cycle Attention Network (DIICAN) is proposed to extract battery multiple states and model battery states' evolving process based on the two intra-cycle and inter-cycle temporal structures. As illustrated in Fig.~\ref{fig:framework}, DIICAN has the three modules: feature extraction module (FEM), temporally structured recurrent module (TRM) and state-coupled regression module (RM).
\begin{figure}[h]
  \centering
  \includegraphics[width=0.75\columnwidth]{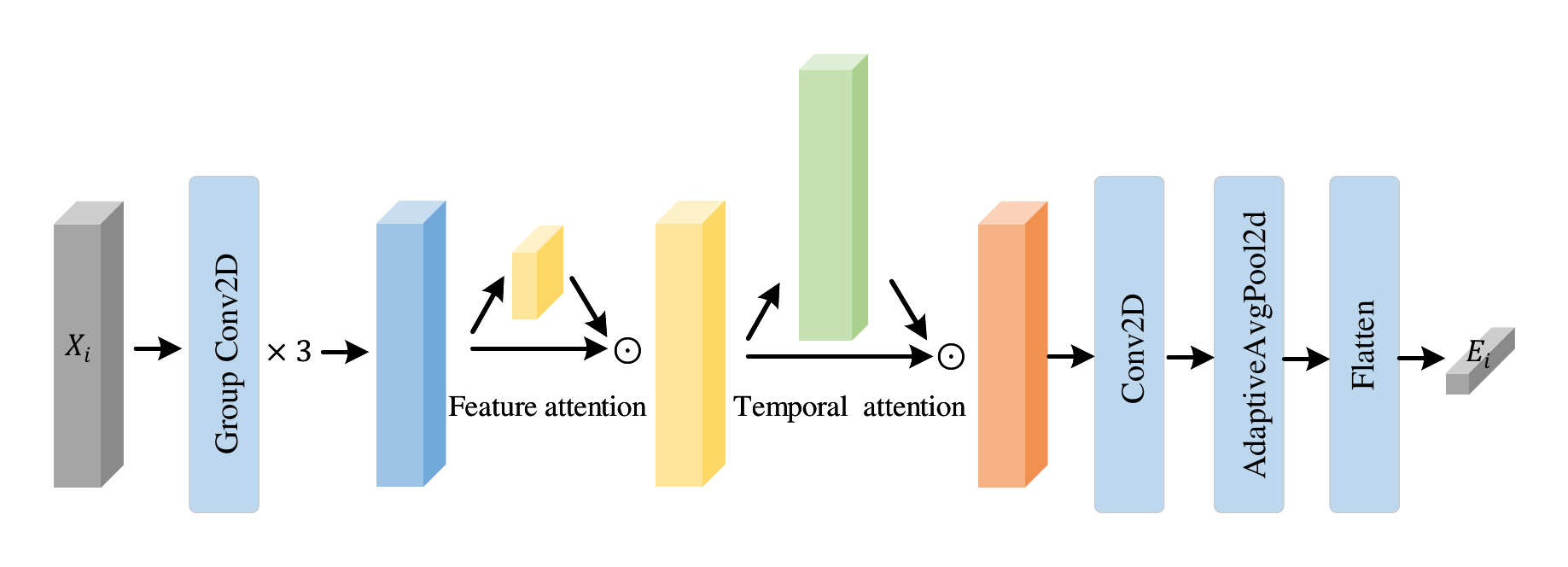}
  \caption{The structure of the feature extraction module.}
  \label{fig:FEM}
\end{figure}

\subsection{Inter and intra-cycle feature representations}
The BMS streaming data is made of the continuous measurement of the battery external electrical and thermal performance that contains the information of the battery states and characterize the battery degradation. For battery states, one is the intra-cycle states such as state of charge (SOC), the other is the inter-cycle states such as state of health (SOH) and remaining of useful life (RUL). Therefore, the BMS streaming time series data can be decomposed into the inter-cycle and intra-cycles features that represent the two temporal structures within the cycles and between the cycles.
\begin{table}[htbp]
  \caption{The inter and intra-cycle feature set.}
  \begin{tabular}{lll}
    \hline
    \textbf{Category} & \textbf{Feature Group} & \textbf{Dimensionality} \\ \hline
    \multirow{4}{*}{inter-cycle  $\mathbf{X}_i$}  & Charging Voltage Curve $(\mathbf{V}^c)$ & $100 $   \\
    & Charging Temperature Curve $(\mathbf{T}^c)$& $100  $ \\
    & Discharging Voltage Curve $(\mathbf{V}^d)$& $100  $ \\
    & Discharging Temperature Curve $(\mathbf{T}^d)$& $100  $ \\ \hline
    \multirow{3}{*}{intra-cycle  $\mathbf{x_j}$}  & Discharging Voltage Point& $1  $ \\ 
    & Discharging Current Point& $1 $ \\
    & Discharging Temperature Point& $1  $ \\ \hline
  \end{tabular}
  \label{tab:fea_set}
\end{table} 

For the SOH and RUL estimation, all the features are embedded in the sequence of charging or discharging curves. The relationship between battery V/I/T curves and battery health status is very difficult to establish in the full battery lifetime. The inter-cycle features, $\mathbf{X}_i=\{X_{i},X_{i+1},\cdots,X_{i+L-1}\}$, are the sequences of voltage, current and temperature (V/I/T) curves in the battery charging/discharging processes measured by BMS within different cycles, which typically include $\mathbf{V}^c=\{V_1^c, V_{2}^c,\cdots, V_N^c\}$, $\mathbf{I}^c=\{I_1^c, I_{2}^c,\cdots, I_N^c\}$, $\mathbf{T}^c=\{T_1^c, T_{2}^c,\cdots, T_N^c\}$, $\mathbf{V}^{d}=\{V_1^d, V_{2}^d,\cdots, V_N^d\}$, $\mathbf{I}^d=\{I_1^d, I_{2}^d,\cdots, I_N^d\}$, and $\mathbf{T}^d=\{T_1^d, T_{2}^d,\cdots, T_N^d\}$ in the $i^{th}$ cycle.  Although the BMS streaming are recorded at the same sampling rate, the total time duration could vary for different batteries and in different cycles. Given that the charged and discharged capacity increases monotonously, the battery V/I/T curves are normalized as the functions of capacity ratio. As a result, the inter-cycle feature $\mathbf{X}_i$ are re-labelled with the capacity indices\cite{yang2021machine} to remove the temporal sampling rate discrepancy. On the other hand, for the SOC estimation, the intra-cycle features, $\mathbf{x}_j$, are the sequences of voltage, current and temperature measurement points (V/I/T) within a discharging cycle. The intra-cycle feature sequence is defined as $\mathbf{x}_j=\{x_{j},x_{j+1},\cdots,x_{j+l}\}$ which $\mathbf{x}_j$ are $l$ historical V/I/T points. The inter-cycle and intra-cycle features for the battery state forecasting in DIICAN are presented in Table~\ref{tab:fea_set}.

\subsection{Feature extraction module}
The feature extraction module is vital to identify the features that can accurately and completely cover the information of each original historical step and the correlation between the attributes. For SOC estimation it applies one fully connected layer (FC) for embedding each feature point $\mathbf{x_j}$ into $\mathbf{e_j}$.  The relationship between battery V/I/T curves and battery health status is very difficult to perceive owing to the complex electrochemical reactions and mechanisms inside the batteries. Thus, SOH-RUL estimation consists of complex structures for the input data $\mathbf{X_i}$ during one cycle to generate the embedding features $\mathbf{E}_i$ associated with battery degradation. The details of the convolution structures are presented below.
\begin{figure}[htbp]
  \centering
  \subfigure[Conv2D]{
  \includegraphics[width=0.2\columnwidth]{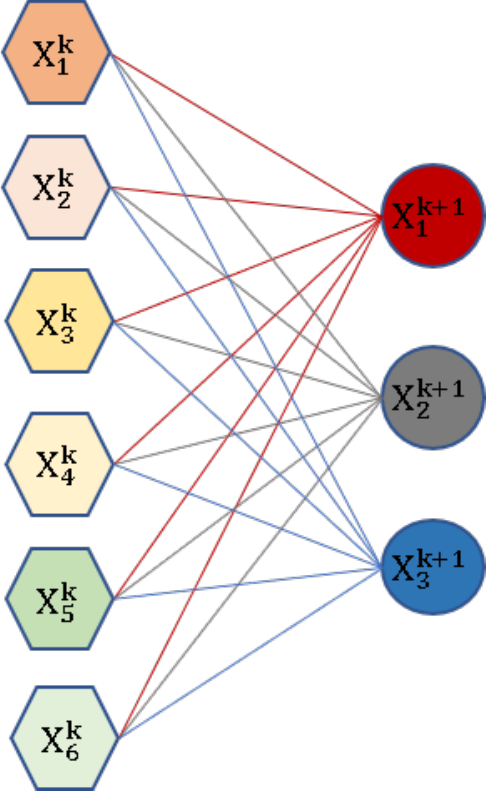}
  \label{fig:2DCNN}}
  \hspace{20mm}
  \subfigure[Group Conv2D]{
  \includegraphics[width=0.2\columnwidth]{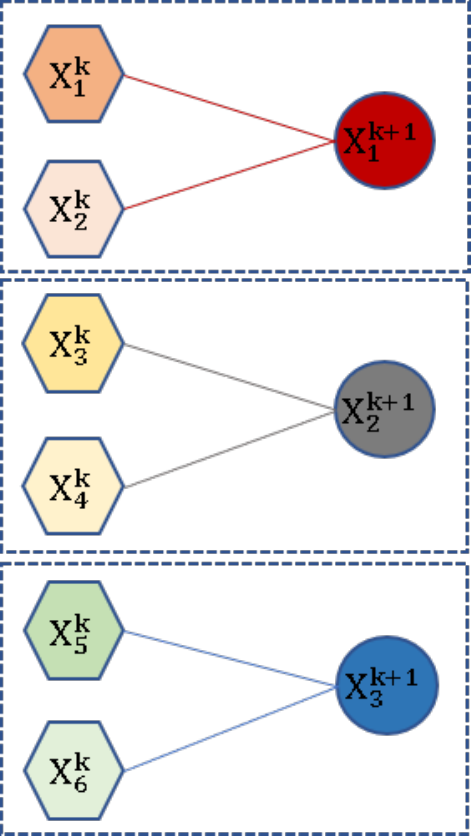}
  \label{fig:g2DCNN}}
  \caption{The Conv2D and group Conv2D.}
  \label{fig:cnn}
\end{figure}

Traditionally, convolutional neural networks (CNNs) have been widely used in the computational vision field. \citet{sun2022excimer} applied the CNN architecture to the time series prediction with excellent performance. The convolution operation is good at capturing the temporal correlation of local information, and the convolution kernel coefficient can flexibly adjust the size of the receptive field to obtain more features of input data in different time scales. Inspired by that, CNNs are utilized to extract degradation-related features hidden in battery V/I/T curves comprehensively and automatically; meanwhile, feature and temporal attention block is adopted to enhance the performance. Each input vector $\mathbf{X_i} \in \mathbb{R} ^{N\times M\times 1 }$ is treated like a color image as the input of a two-dimensional convolutional neural network. Fig.~\ref{fig:FEM} indicates the complex 2DCNN structure for SOH-RUL estimation.
\begin{figure}[htbp]
  \centering
  \includegraphics[width=0.7\textwidth]{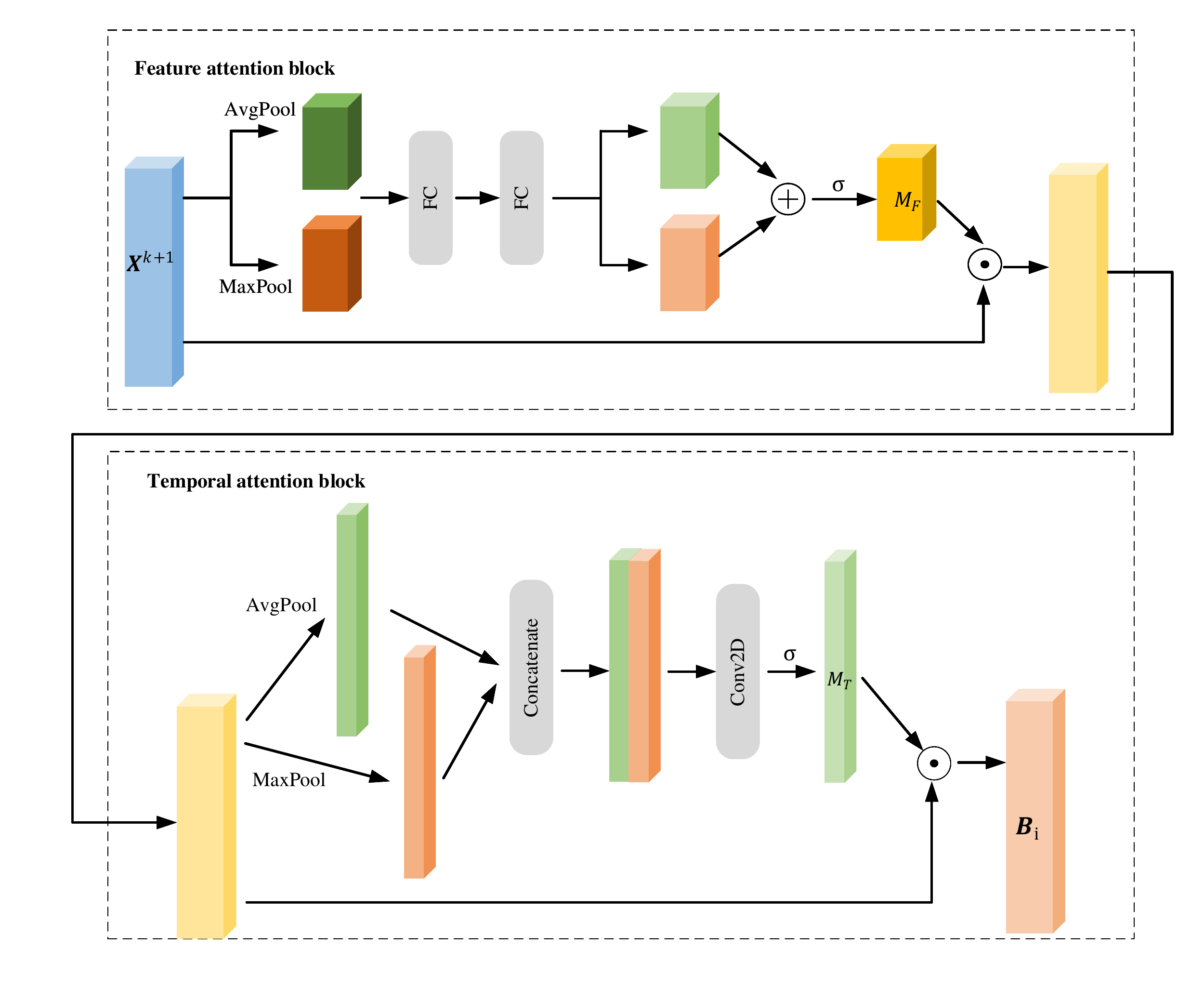} 
  \caption{The feature and temporal attention maps.}
  \label{fig:f-t-att}
\end{figure}

The convolution layer is computed as convolving the input feature maps with filters. Taking the $k^{th}$ layer for an example, the input of the $k^{th}$ layer can be denoted as  $\mathbf{X}^k = \{X_1^k, X_2^k, ... , X_{N}^k\}$, where $N$ is the input channel number and $X_s^k$ is the $s^{th}$ feature map. The filters of the $k^{th}$ layer are denoted as $\mathbf{W}^k=\{W_1^k, W_2^k, ... , W_O^k\}$, where $O$ denotes the filter number, as well as the output channel number, and $W_s^k$ is the $s^{th}$ 2D convolutional filter.  The Conv2D structure is defined as,  
\begin{align}
  \mathbf{X}^{k+1} & = \mathbf{W}^k \otimes  \mathbf{X}^k \\
          & = \{W_1^k \ast \mathbf{X}^{k}, W_2^k \ast \mathbf{X}^{k}, ... , W_{O}^k \ast \mathbf{X}^{k}\}
\end{align} where $\otimes$ denotes the convolution between two sets, $\ast$ denotes the convolution operation between a filter and the input feature maps. After each Conv2D layer, the rectified linear unit (ReLU) activation function $f(x) = \max(0, x)$ is applied. 

As shown in Fig.~ \ref{fig:g2DCNN}, the group convolution is a special case of a sparsely connected convolution \cite[]{huang2018condensenet}. In group convolution, the input feature maps $\mathbf{X}^k$ are divided into $G$ groups equally as the number of filters, {\it{i.e.}}, $\mathbf{X}^k = \{\mathbf{X}_1^k, \mathbf{X}_2^k, ... , \mathbf{X}_G^k\}$\unboldmath and $W^k=\{W_1^k, W_2^k, ... ,W_G^k\}$.  $\mathbf{X}^{k+1}$ is reformulated as below,
\begin{equation}
  \mathbf{X}^{k+1} = \{W_1^k \otimes \mathbf{X}_1^{k}, W_2^k \otimes \mathbf{X}_2^{k}, ... , W_{G}^k \otimes \mathbf{X}_{G}^{k}\}.
\end{equation} The group Conv2D reduces the computational cost by partitioning the input features into $G$ mutually exclusive groups producing its own output feature maps. The computational cost is reduced by a factor $G$ to be $\frac{O\times N}{G}$.Each group represents one feature, which means that each convolution is operated on one feature. 
\begin{figure}[htbp]
  \centering
  \includegraphics[width=0.8\columnwidth]{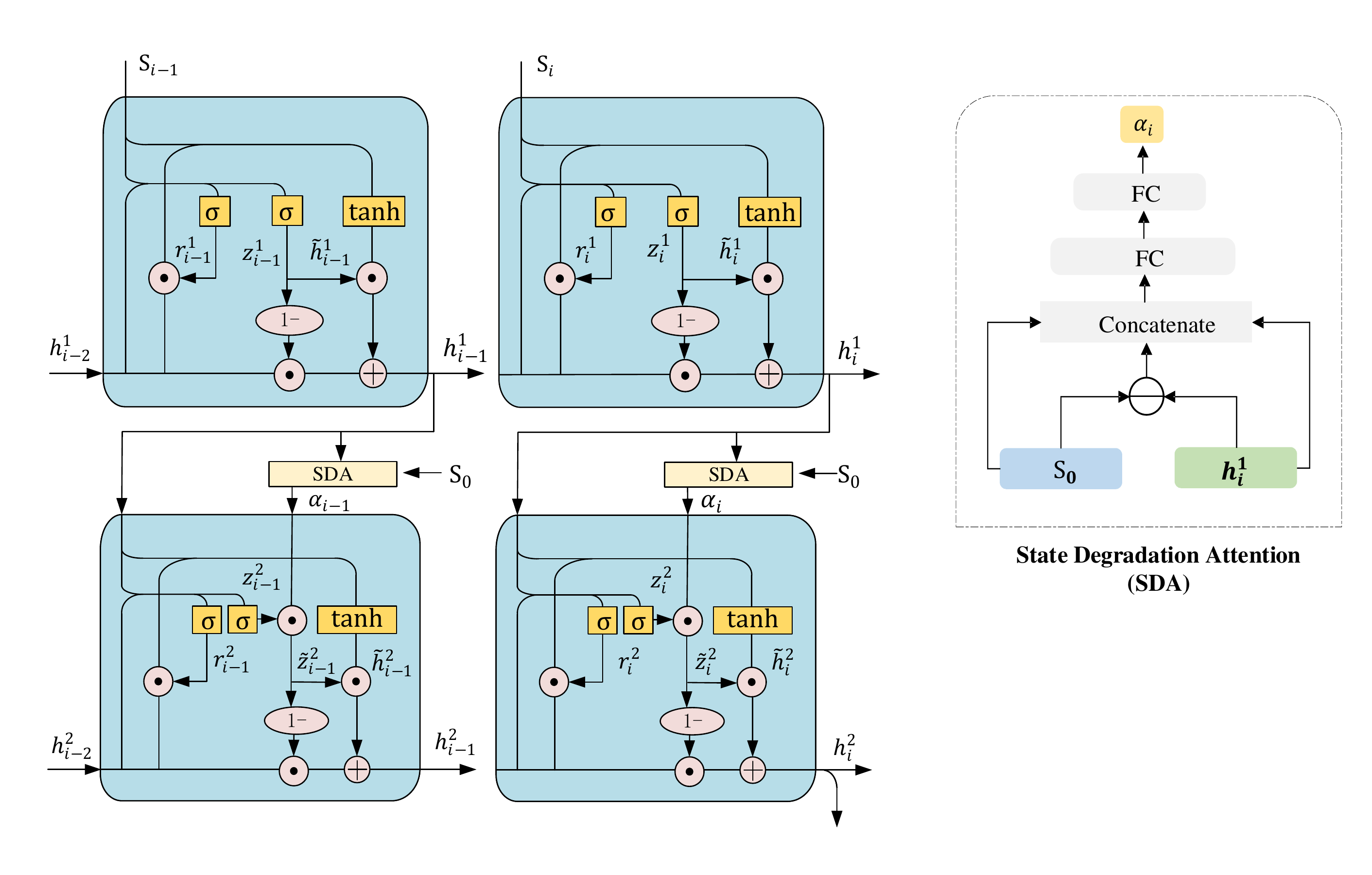} 
  \caption{The structure of the temporally structured recurrent module.}
  \label{fig:2gru}
\end{figure}

With the feature map $\mathbf{X}^K$ transformed by three group Conv2D layers, a feature/temporal attention map $\mathbf{B}_i$ are generated with the feature attention block $\mathbb{M}_F$ and temporal attention block $\mathbb{M}_T$.  As shown in Fig.~\ref{fig:f-t-att}, the feature/temporal attention map $\mathbf{B}_i$ is defined as,
\begin{equation}
    \mathbf{B}_i= \mathbb{M}_T(\mathbb{M}_F(\mathbf{X}_i^{K}) \odot \mathbf{X}_i^{K}) \odot (\mathbb{M}_F(\mathbf{X}_i^{K}) \odot \mathbf{X}_i^{K}) 
\end{equation}
where $\odot$ denotes the element-wise multiplication. During multiplication, attention values are broadcasted accordingly.

A feature attention map is inferred by exploiting the inter-feature relationship. To weight the relevance of features, the feature attention block $\mathbb{M}_F$ is defined as,
\begin{equation}
  \mathbb{M}_F (\bullet)  = \sigma(MLP(AvgPool(\bullet)) + MLP(MaxPool(\bullet)))
\end{equation}
where $\sigma$ denotes the sigmoid function and MLP denotes the multilayer perceptron. To compute the feature attention efficiently, feature dimension is squeezed. By aggregating features with both average-pooling and max-pooling operations simultaneously, the feature attention block ends up with MLP and sigmoid function. 

Furthermore, temporal attention is computed along the time dimension. As shown in Fig.~\ref{fig:f-t-att}, the temporal attention block $\mathbb{M}_T$ is defined as:
\begin{equation}
  \mathbb{M}_T (\bullet) = \sigma(\mathcal{F}^{5\times1} (AvgPool(\bullet)\parallel MaxPool(\bullet)) 
\end{equation}
where $\sigma$ denotes the sigmoid function, $\parallel$ is a concatenation operation and $\mathcal{F}^{5\times1}$ represents a convolution operation with the $5\times1$ kernel filter. To compute the temporal attention maps, this paper applies the average-pooling and max-pooling operations along the feature dimension and concatenates them to generate efficient feature descriptors $AvgPool(\bullet) $ and $MaxPool(\bullet)$. Applying pooling operations along the temporal dimension is effective in highlighting informative temporal regions. With the concatenated feature descriptor, a temporal attention map $\mathbf{M}_{T}$ is generated through one convolution layer.  

Finally, the dimension reshaping block consists of one Conv2D layer with a $1\times 1$ kernel filter and one adaptive 2D average pooling layer. The feature/temporal attention map $\mathbf{B}_i$ is flattened into the inter-cycle embedding features $\mathbf{E}_i$. 

\subsection{Temporally structured recurrent module}
\label{subsubsec:sem}
The temporally structured recurrent module is developed to extract battery state evolution pattern. As illustrated in Fig.~\ref{fig:2gru}, it has the many-to-one structure with the gated recurrent unit (GRU), GRU with attentional gate (AUGRU) and the state degradation attention (SDA) unit. It takes the states $\mathbf{S}_i$ as input which can be either the inter-cycle embedding features $\mathbf{E}_{i}$ or intra-cycle embedding features $\mathbf{e}_{j}$.
\begin{figure}[htbp]
  \centering
  \includegraphics[width=0.4\columnwidth]{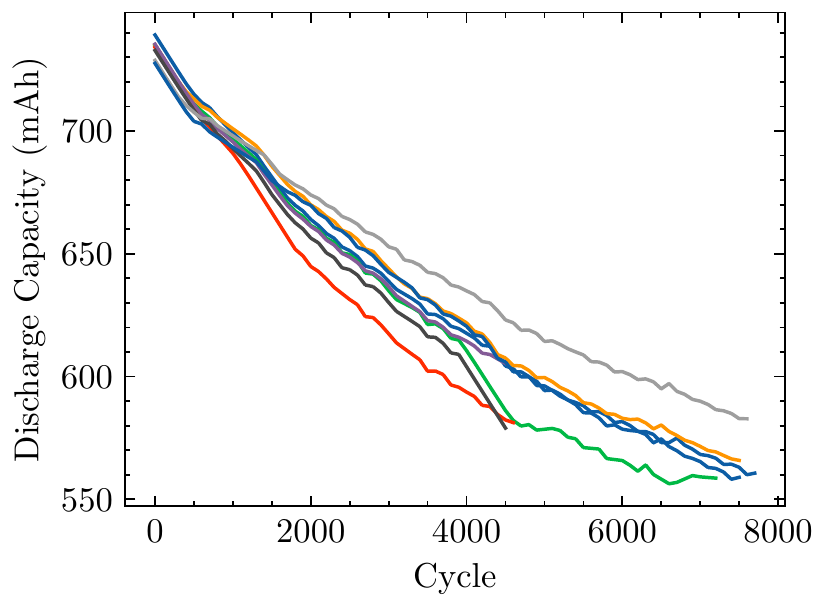} 
  \caption{The capacity degradation curves of eight cells in the Oxford dataset.}
  \label{fig:capacity}
\end{figure}

The sequence of the inter-cycle and intra-cycle embedding features $\mathbf{S}_{i-d-1},\mathbf{S}_{i-d},\cdots,\mathbf{S}_{i}$ are concatenated and fed into GRU, and then the sequence of hidden states $\mathbf{h}_{i-d-1}^1,\mathbf{h}_{i-d}^1,\cdots, \mathbf{h}_{i}^1$ from GRU and the $\mathbf{S}_0$ of the initial battery embedding features are fed into AUGRU to measure how much the battery states deviate from the initial battery states. With two gates, reset gate $\mathbf{r}_i^1$ and update gate $\mathbf{z}_i^1$, the GRU is defined as, 
\begin{align}
  \mathbf{r}_i^1  &= \sigma (\mathbf{W}_{rh}^1 \mathbf{h}_{i-1}^1 + \mathbf{W}_{rn}^1 \mathbf{S}_i + \mathbf{b}_r^1) \notag \\
  \mathbf{z}_i^1  &= \sigma (\mathbf{W}_{zh}^1 \mathbf{h}_{i-1}^1 + \mathbf{W}_{zn}^1 \mathbf{S}_i + \mathbf{b}_z^1) \notag \\
  \tilde{\mathbf{h}}_i^1  &= \tanh (\mathbf{W}_{hh}^1 \mathbf{h}_{i-1}^1 + \mathbf{W}_{hn}^1 \mathbf{S}_i + \mathbf{b}_h^1) \notag \\
  \mathbf{h}_i^1 &= (1-\mathbf{z}_i^1)\odot \tilde{\mathbf{h}}_i^1 + \mathbf{z}_i^1\odot \mathbf{h}_{i-1}^1
\end{align}
where the superscript $1$ denotes the GRU, ($\mathbf{W}_{rh}^1$, $\mathbf{W}_{zh}^1$, $\mathbf{W}_{hh}^1$), ($\mathbf{W}_{rn}^1$, $\mathbf{W}_{zn}^1$, $\mathbf{W}_{hn}^1$), and ($\mathbf{b}_r^1$, $\mathbf{b}_z^1$, $\mathbf{b}_h^1$) denote the weight, $\tilde{\mathbf{h}}_i^1$ represents the hidden state, and $\mathbf{h}_{i-1}^1$ and $\mathbf{h}_{i}^1$  denote the hidden states in and out of the GRU, respectively.

Inspired by the biological systems of humans that tend to focus on the distinctive parts when processing large amounts of information, \citet{firat2016multi}, the attention mechanism is used to improve the efficiency and accuracy of perceptual information processing. In order to extract battery state evolution pattern by measuring the deviation of the embedding feature $\mathbf{S}_i$ at $i^{th}$ from the embedding feature $\mathbf{S}_0$ at the initial cycle, the initial battery embedding features $\mathbf{S}_{0}$ and hidden states $\mathbf{h}_{i}^1$ are fed into the state degradation attention (SDA) unit.

As demonstrated in Fig.~\ref{fig:2gru}, the SDA consists of two fully connected layers. $\mathbf{h}_{i}^1$, $\mathbf{S}_{0}$ and $\mathbf{h}_{i}^1\ominus \mathbf{S}_{0}$ are concatenated and fed into the SDA to compute the attention weight associated with the battery degradation in Eq.~\ref{eqa:att_normal},
\begin{align}
 \alpha_{i} = att([\mathbf{h}_{i}^1 \parallel \mathbf{S}_{0} \parallel \mathbf{h}_{i}^1\ominus \mathbf{S}_{0}]) 
 \label{eqa:att_normal}
\end{align}
where $\parallel$ is a concatenation operation, $\ominus$ denotes element-wise minus and $att$ represents the state degradation attention unit. Aiming to accurately account for the battery degradation degree, we use the absolute attention weights instead of the relative attention weight distribution and the softmax normalization of attention weights is abandoned. 
\begin{figure}[htbp]
  \centering
  \includegraphics[width=0.4\columnwidth]{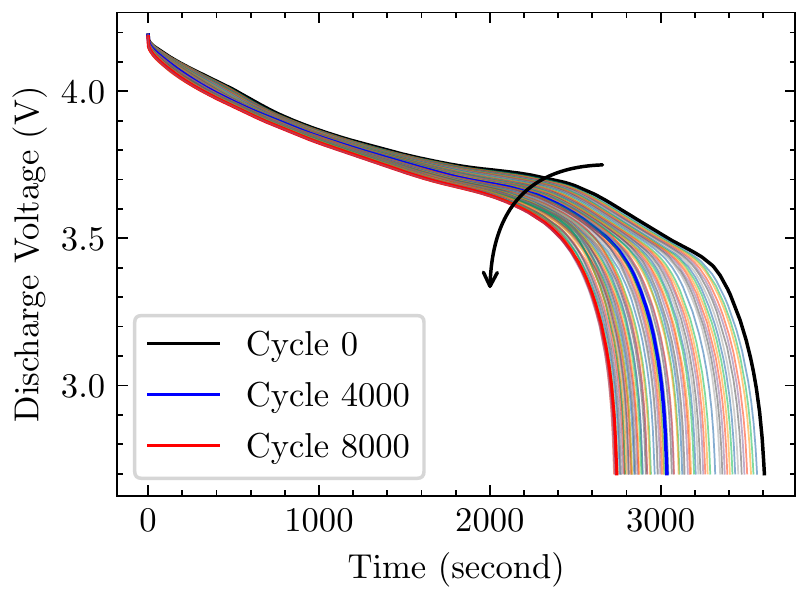} 
  \caption{The discharge voltage curves of all the cycles in the full lifespan for Cell 1 in the Oxford dataset.}
  \label{fig:vd_t}
\end{figure}

Then, AUGRU\cite[]{zhou2019deep} is used to calculate the hidden states. As shown in Fig.~\ref{fig:2gru}, it combines the attention mechanism and the GRU update gates together to learn the battery state deviation, which is defined as,
\begin{align}
\mathbf{r}_i^2 &= \sigma (\mathbf{W}_{rh}^2 \mathbf{h}_{i-1}^2 + \mathbf{W}_{rn}^2 \mathbf{h}_i^1 + \mathbf{b}_r^2) \notag \\
 \mathbf{z}_i^2 &= \sigma (\mathbf{W}_{zh}^2 \mathbf{h}_{i-1}^2 + \mathbf{W}_{zn}^2 \mathbf{h}_i^1 + \mathbf{b}_z^2) \notag \\
 \tilde{\mathbf{h}}_i^2 &= \tanh (\mathbf{W}_{hh}^2 \mathbf{h}_{i-1}^2 + \mathbf{W}_{hn}^2 \mathbf{h}_i^1 + \mathbf{b}_h^2) \notag \\
 \tilde{\mathbf{z}}_i^2 &= \alpha_{i} \ast \mathbf{z}_i^2 \notag \\
 \mathbf{h}_i^2 &= (1-\tilde{\mathbf{z}}_i^2)\odot \tilde{\mathbf{h}}_i^2 + \tilde{\mathbf{z}}_i^2\odot \mathbf{h}_{i-1}^2
\end{align}
where the superscript $2$ represents the AUGRU, $\ast$ means scalar-vector product
$\mathbf{z}_i^2$ is the update status, $\tilde{\mathbf{z}}_i^2$ indicates the attention update gate in the AUGRU, and $\mathbf{h}_i^2$, $\tilde{\mathbf{h}}_i^2$, and $\mathbf{h}_{i-1}^2$ indicate the hidden states of AUGRU. The attention weight is added as the attentional update gate which keeps original dimensional information of update gate, and decides the importance of each dimension. AUGRU models the battery state degradation smoothly. The last hidden state $\mathbf{h}_i^2$ from the AUGRU is fed into the state-coupled regression module.
\begin{figure}[htbp]
  \centering
  \includegraphics[width=0.4\columnwidth]{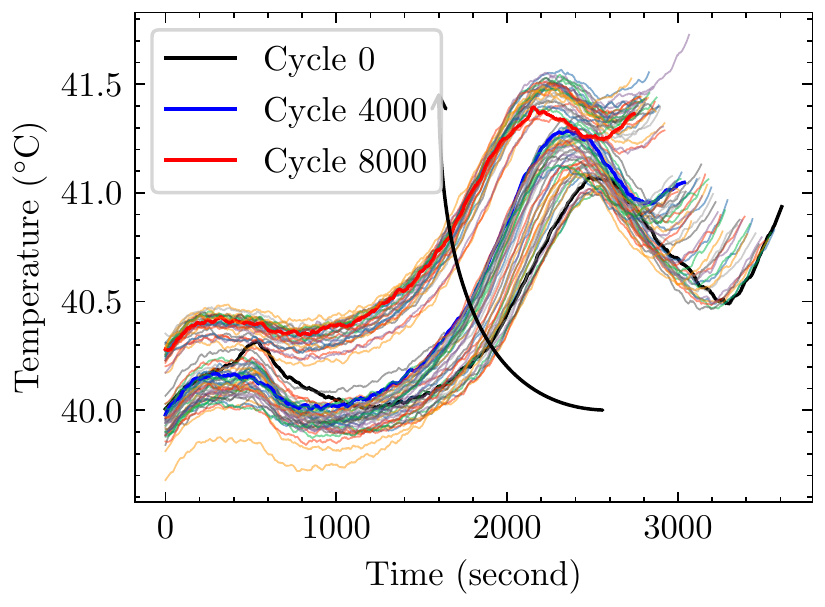} 
  \caption{The discharge temperature curves of all the cycles in the full lifespan for Cell 1 in the Oxford dataset.}
\end{figure}

\subsection{state-coupled regression module}
In the end, the regression module is composed of two fully connected layers for the inter-cycle SOH/RUL estimator and intra-cycle SOC estimator. Due to the nature of battery degradation dynamics, the coupling of SOC and SOH should be considered. 
\begin{figure}[htbp]
  \centering
  \label{fig:SOC_t}
  \includegraphics[width=0.4\columnwidth]{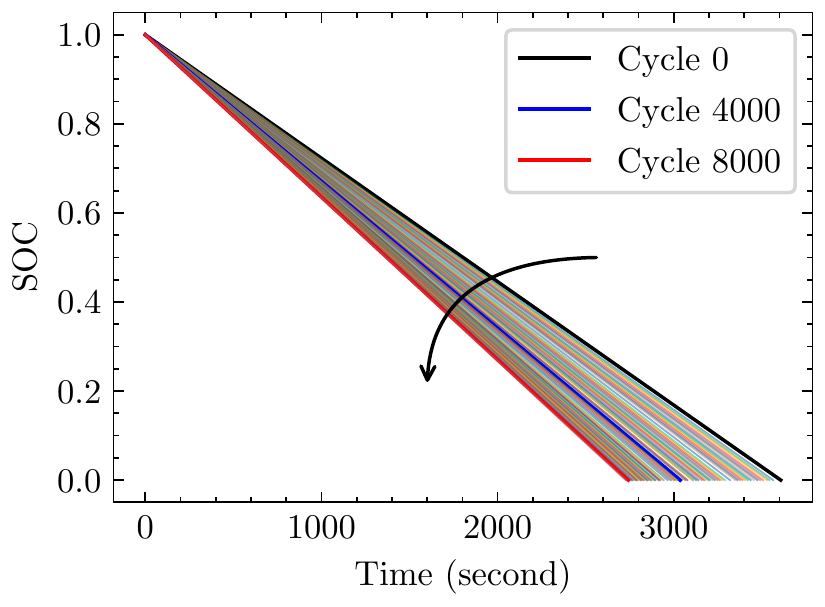} 
  \caption{The SOC curves of of all the cycles in the full lifespan for Cell 1 in the Oxford dataset.}
  \label{fig:T_t}
\end{figure} 

In the SOH/RUL estimator, the two parallel MLP layers are applied for the SOH and RUL estimation. The inter-cycle hidden states are fed into the two MLP layers. And each task should set a loss function, and then all the loss functions are weighted and accumulated to form an overall loss function. The network model is trained by gradient back propagation to optimize the overall loss function. The definition of loss function is as follows: 
\begin{equation}
  Loss = (1-\beta)\times loss_{{RUL}} + \beta \times loss_{{SOH}}
\end{equation}
where $loss_{RUL}$ and $loss_{SOH}$ represent the loss functions for the RUL and SOH estimation respectively, and $\beta$ denotes the loss weight. 

For the SOC estimator, both intra-cycle and inter-cycle hidden states are fed into the MLP layer for the SOC estimator. To fully consider the aging characteristics, inter-cycle hidden states as battery degradation-related state is coupled in SOC estimation for capacity calibration to improve estimation performance over the whole lifespan.  In order to avoid curse of dimensionality, the inter-cycle hidden states are converted trough one fully connected layer and then concatenated together with the intra-cycle hidden states to estimate SOC.

\section{Battery dataset and DIICAN training}\label{sec:training}
\subsection{Dataset}
\label{sec:dataset}
The Oxford battery degradation dataset\cite[]{birkl2017oxford} is used to evaluate the performance of the proposed DIICAN method. The Oxford battery aging dataset have long term battery ageing tests. The tests consist of aging experiments applied to eight commercial Kokam pouch cells of 740-mAh nominal capacity, numbered as Cell1 - Cell8 respectively. The cells were exposed to a 1C or C/25 constant-current-constant-voltage charging profile, followed by a driving cycle discharging profile obtained from the Urban Artemis profile.  After every 100 cycles, a characterization test was carried out including a full charge-discharge cycle at 1C. The discharge capacity degradation curves of eight cells are shown in Fig.~\ref{fig:capacity}. It's obvious that capacity degradation curves have the same downward trend, which is consistent with the real-world decay process of lithium-ion batteries. 
\begin{table*}[htbp]
  \centering
  \caption{The hyper-parameters of the proposed DIICAN method.}
  \resizebox{\textwidth}{!}{
  \begin{tabular}{lllll}
    \hline
    \multirow{2}{*}{\textbf{Module}} & \multicolumn{2}{l}{\textbf{Inter-cycle SOH-RUL Estimation}} & \multicolumn{2}{l}{\textbf{Intra-cycle SOC Estimation}}\\ \cline{2-5}
    & \textbf{Structure} & \makecell[l]{\textbf{Details} (channel number/\\kernel size/stride, in\_out\_sizes)} & \textbf{Structure} & \textbf{Details}(in\_out\_sizes) \\
    \hline
    \multirow{9}{*}{\textbf{FEM}} & Group Conv2d, ReLU & (4, 128)/(5, 1)/(1, 1)& \multirow{9}{*}{FC, ReLU}&\multirow{9}{*}{(3, 32)} \\
    & Group Conv2d, ReLU & (128, 128)/(5, 1)/(1, 1)& &  \\ 
    & Group Conv2d, ReLU & (128, 128)/(5, 1)/(2, 2)& & \\ 
    
    & AdaptiveAvgPool2d; AdaptiveMaxPool2d & (-, 1); (-, 1) & & \\ 
    & Conv2d, ReLU &(128, 8)/(1, 1)/(1, 1)& & \\ 
    & Conv2d, Sigmoid &(8, 128)/(1, 1)/(1, 1)& & \\ 

    & Conv2d, Sigmoid & (2, 1)/(5, 5)/(1, 1)/(2, 2)& & \\ 

    &Conv2d &  (128, 8)/(1, 1)/(1, 1) & & \\ 
    &AdaptiveAvgPool2d,  ReLU&  (-, 1) & & \\ \hline

    \multirow{5}{*}{\textbf{TRM}}& GRU & (128, 128)& GRU& (32, 32) \\ 
    & FC, ReLU  &(384, 128)& FC, ReLU&(96, 32) \\ 
    & FC, ReLU  &(128, 32)&FC, ReLU &(32, 8) \\ 
    & FC  &(32, 1)& FC  &(8, 1) \\ 
    & AUGRU & (128, 128)& AUGRU & (32, 32) \\ \hline

    \multirow{3}{*}{\textbf{RM}}& & & FC, ReLU& (128, 32) \\
    &FC, ReLU; FC, ReLU &(128, 32); (128, 32)& FC, ReLU& (64, 8) \\ 
    &FC; FC &(32, 1); (32, 1)& FC& (8, 1)\\ \hline
  \end{tabular}}
  \label{tab:para}
\end{table*}

\subsection{Battery states and feature representation}
The state of health (SOH), remaining useful life (RUL) and  state of charge (SOC) are key metrics for the battery management systems at inter and intra-cycle time-scale. These states are determined by the internal physics and chemistry properties and revealed by the external electrical and thermal performance. Thus, the internal parameters of battery can be estimated based on the measurement of the external parameters. 
\begin{figure}[htbp]
  \centering
  \subfigure[Cell 1]{
  \label{fig:soh:cell1}
  \centering
  \includegraphics[width=0.35\columnwidth]{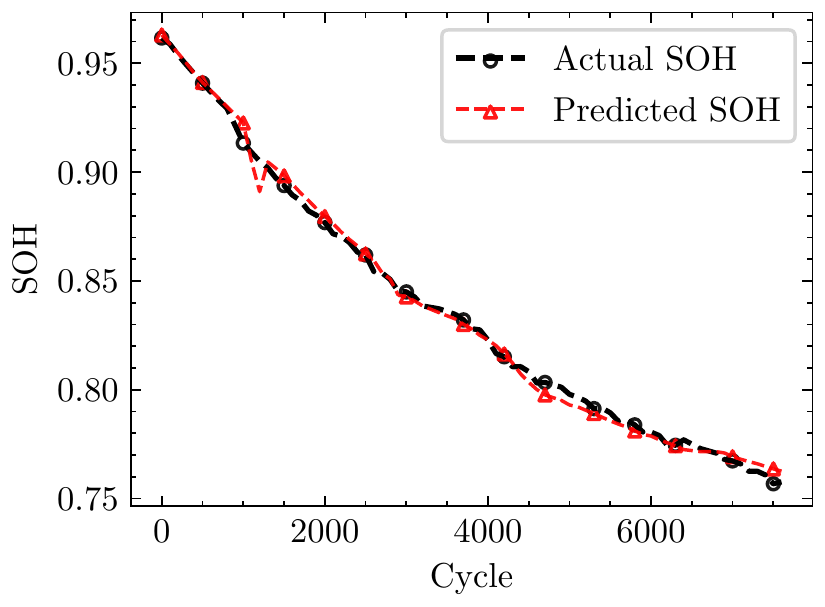} 
  } 
  \subfigure[Cell 2]{
  \label{fig:soh:cell2}
  \centering
  \includegraphics[width=0.35\columnwidth]{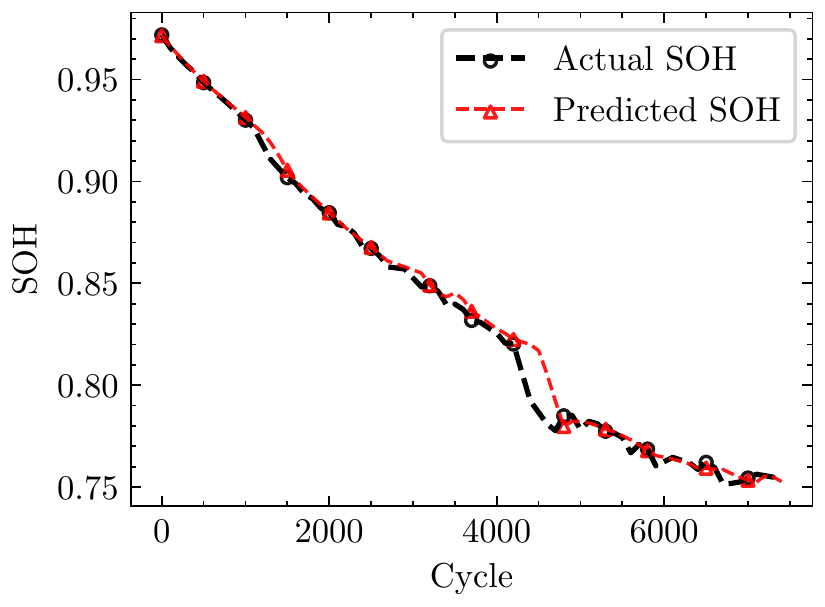} 
  } 
  \subfigure[Cell 3]{
  \label{fig:soh:cell3}
  \centering
  \includegraphics[width=0.35\columnwidth]{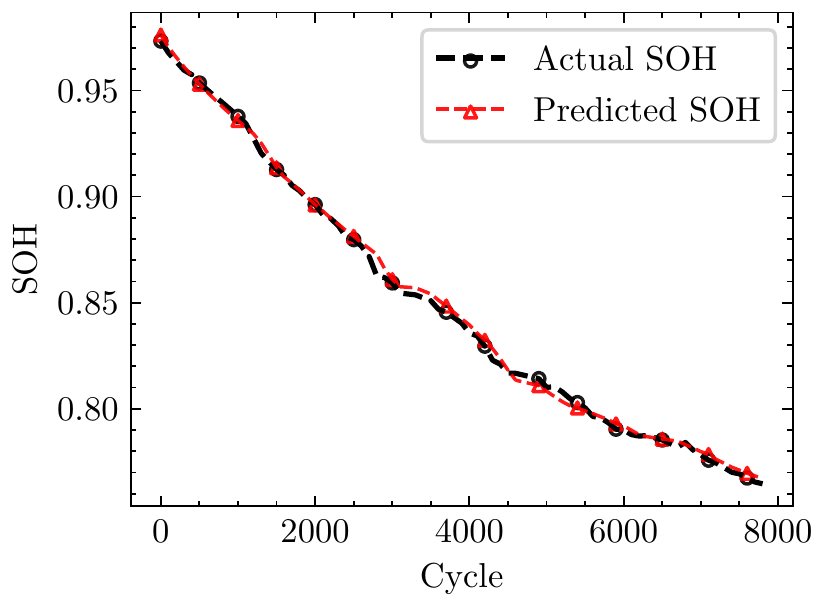} 
  } 
  \subfigure[Cell 4]{
  \label{fig:soh:cell4}
  \centering
  \includegraphics[width=0.35\columnwidth]{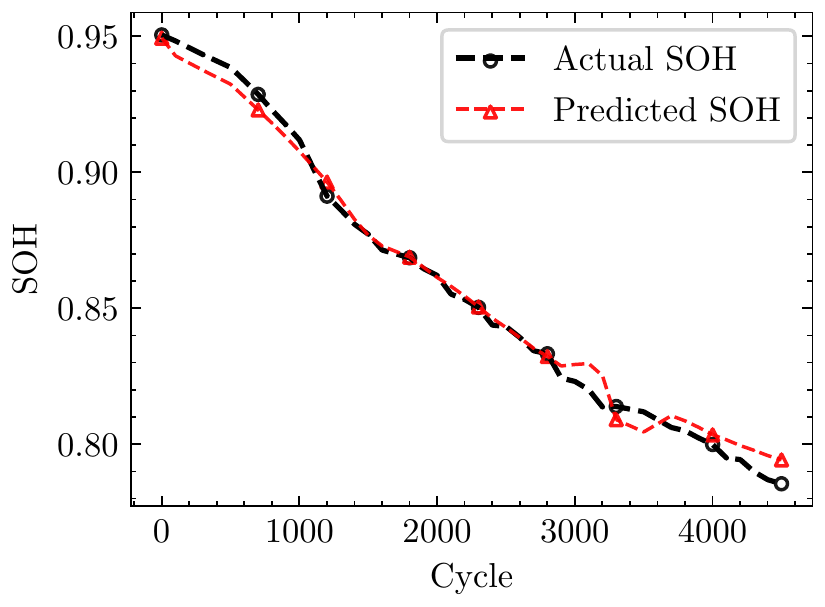} 
  } 
  \subfigure[Cell 5]{
  \label{fig:soh:cell5}
  \centering
  \includegraphics[width=0.35\columnwidth]{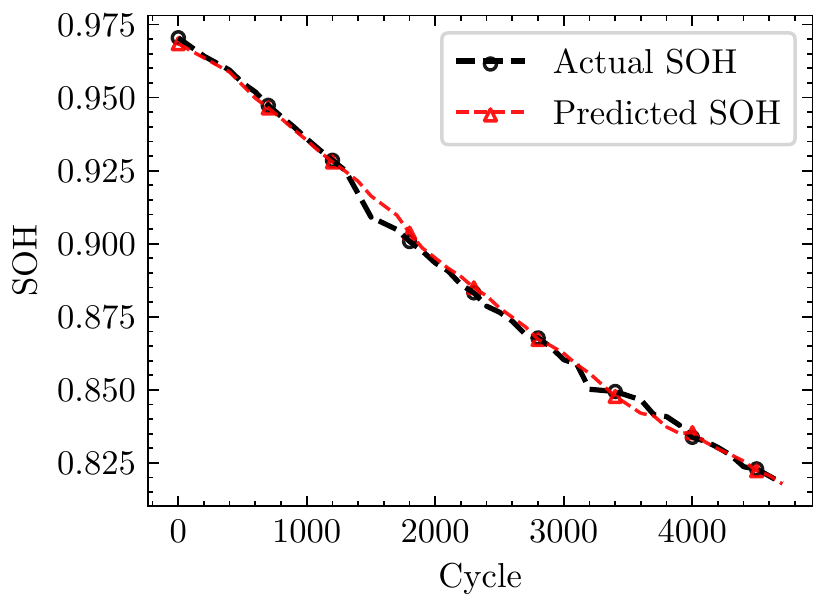} 
  } 
  \subfigure[Cell 6]{
  \label{fig:soh:cell6}
  \centering
  \includegraphics[width=0.35\columnwidth]{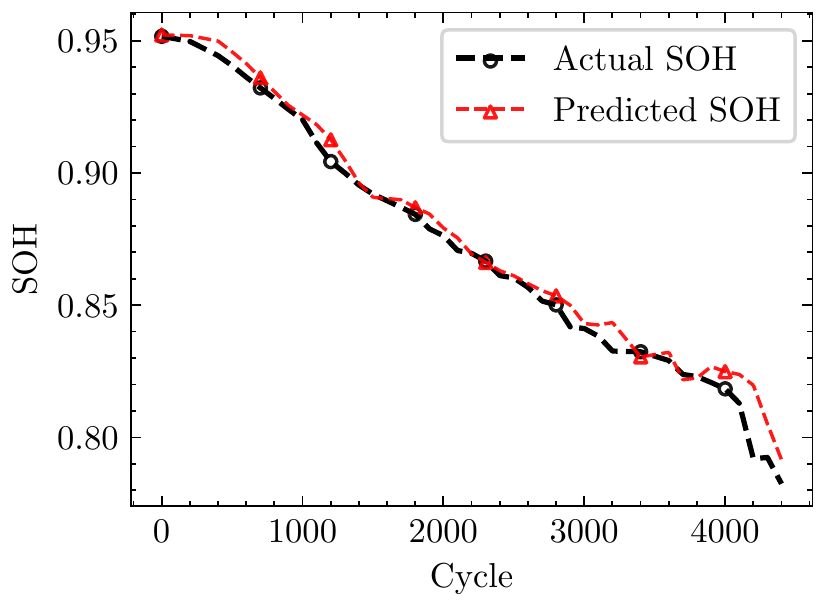} 
  } 
  \subfigure[Cell 7]{
  \label{fig:soh:cell7}
  \centering
  \includegraphics[width=0.35\columnwidth]{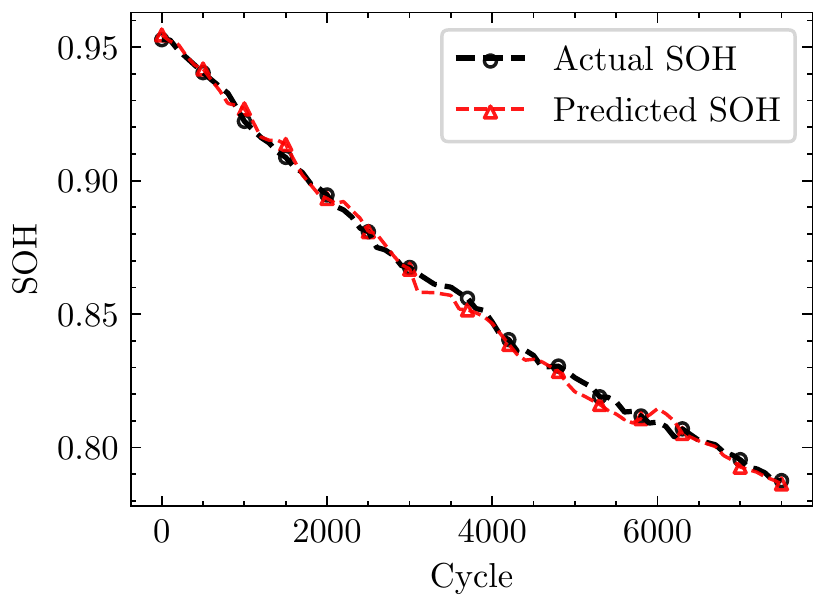} 
  } 
  \subfigure[Cell 8]{
  \label{fig:soh:cell8}
  \centering
  \includegraphics[width=0.35\columnwidth]{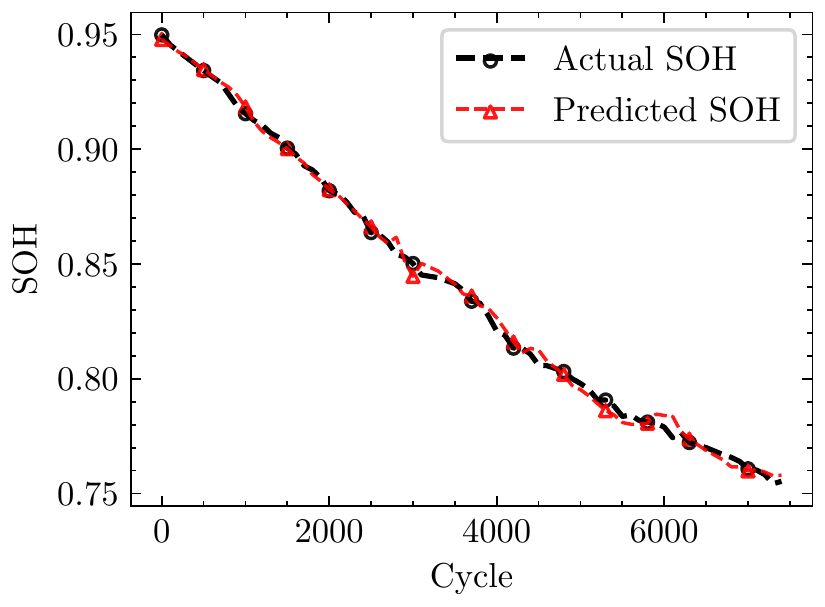} 
  } 
  \caption{The SOH estimation results of the eight cells in the Oxford dataset. The black circle dashed lines are actual SOHs, and the red triangle dashed line predicted SOH with the DIICAN method.}
  \label{fig:soh_ox}
\end{figure}

In Fig.~\ref{fig:vd_t}, the discharge voltage curves in different cycles of a typical cell, cell 1 in the Oxford dataset are shown. With the battery aging, the middle part of the discharge voltage curve gradually become more and more steeper and the derivative of the battery discharge voltage curve becomes larger with the cycle increasing. In other words, the discharge voltage curves are able to reflect the aging process of the battery. Fig.~\ref{fig:T_t} shows the temperature curves in the discharging process. The temperature curves gradually increase with the battery cycles, indicating that the internal impedance of the battery gradually increase. The battery degradation can be measured by analyzing the temperature curves. Fig.~\ref{fig:SOC_t} shows the intra-cycle SOC curves. With the battery aging, the slope of the SOC curve becomes more and more steeper.
\begin{table}[htbp]
  \centering
  \caption{The SOH estimation of the eight cells in the Oxford dataset.}
  \setlength{\tabcolsep}{1.8mm}{
  \begin{tabular}{lcccc}
      \hline
      \textbf{Test Cell} & \textbf{MAE(\%)} & \textbf{MAPE(\%)} & \textbf{RMSE(\%)} & \textbf{$R^2$} \\
      \hline
      Cell 1 &0.277  & 0.0152 & 0.360 & 0.997\\ 
      Cell 2 &0.349  & 0.336 & 0.666 & 0.991\\ 
      Cell 3 &0.221  & 0.138 & 0.268 & 0.998\\ 
      Cell 4 &0.388  & 0.162 & 0.483 & 0.992\\ 
      Cell 5 &0.173  & 0.060 & 0.237 & 0.997\\ 
      Cell 6 &0.467  & 0.514 & 0.672 & 0.979\\ 
      Cell 7 &0.239  & 0.0650 & 0.298 & 0.996\\ 
      Cell 8 &0.227  & 0.0304 & 0.294 & 0.996\\ 
      \hline 
    \end{tabular}} 
    \label{tab:soh_ox}
\end{table}

Based on the above analysis, it should be noted that SOC and SOH are highly coupled due to the electrochemical processes. Therefore, to fully consider the battery aging features of the battery, the hidden states extracted from the inter-cycle features should be used as an input for the intra-cycle SOC estimation.
\begin{figure}[htbp]
  \centering
  \subfigure[Cell 1]{
  \label{fig:rul:cell1}
  \centering
  \includegraphics[width=0.35\columnwidth]{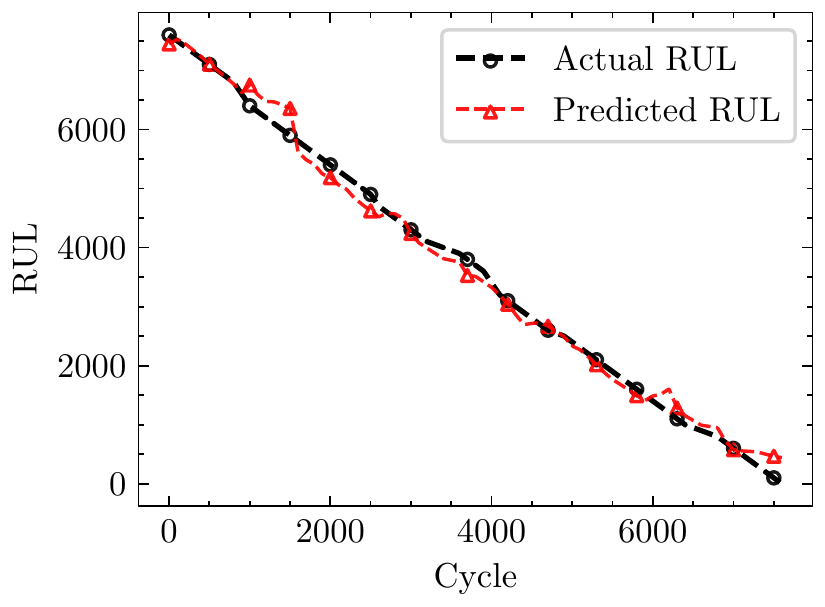} 
  } 
  \subfigure[Cell 2]{
  \label{fig:rul:cell2}
  \centering
  \includegraphics[width=0.35\columnwidth]{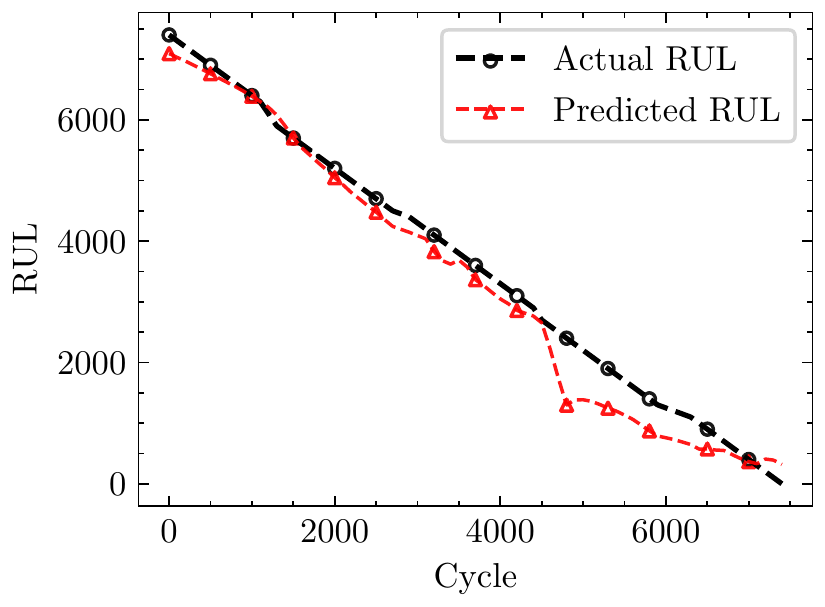} 
  } 
  \subfigure[Cell 3]{
  \label{fig:rul:cell3}
  \centering
  \includegraphics[width=0.35\columnwidth]{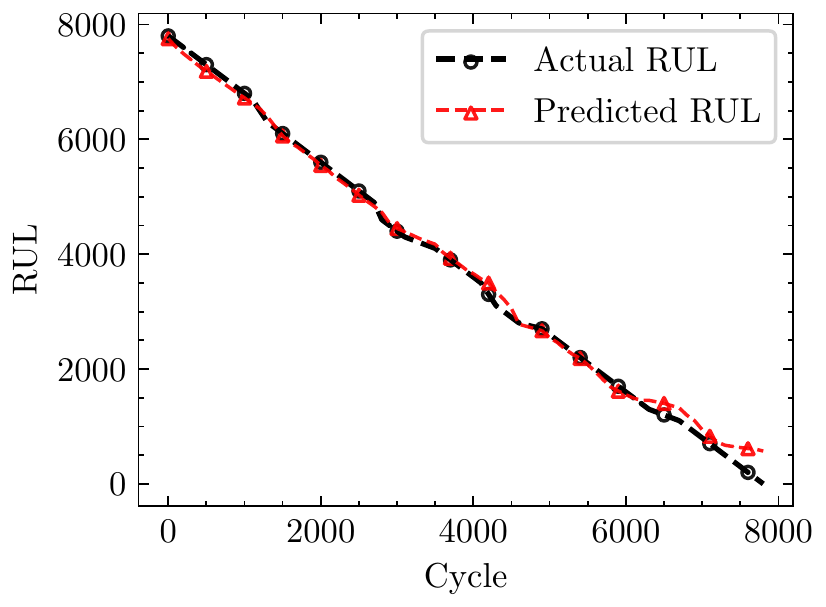} 
  } 
  \subfigure[Cell 4]{
  \label{fig:rul:cell4}
  \centering
  \includegraphics[width=0.35\columnwidth]{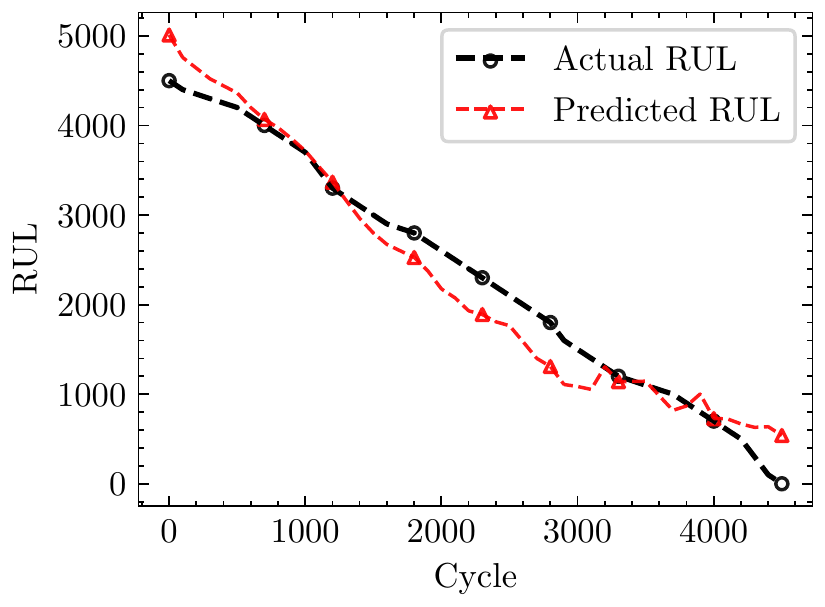} 
  } 
  \subfigure[Cell 5]{
  \label{fig:rul:cell5}
  \centering
  \includegraphics[width=0.35\columnwidth]{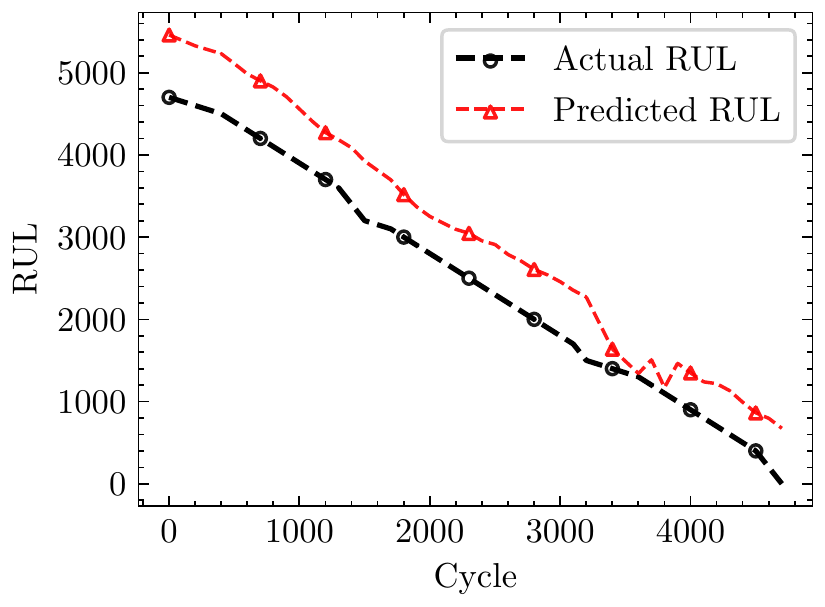} 
  } 
  \subfigure[Cell 6]{
  \label{fig:rul:cell6}
  \centering
  \includegraphics[width=0.35\columnwidth]{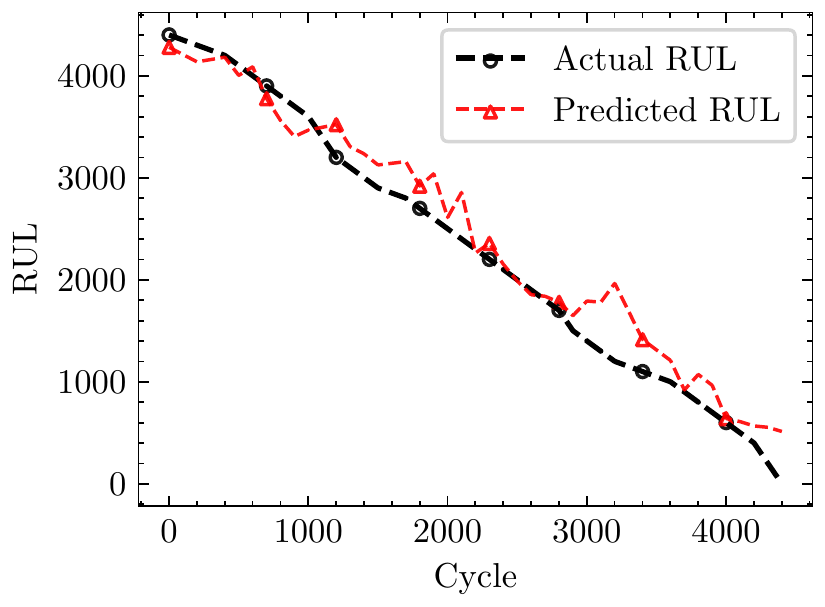} 
  } 
  \subfigure[Cell 7]{
  \label{fig:rul:cell7}
  \centering
  \includegraphics[width=0.35\columnwidth]{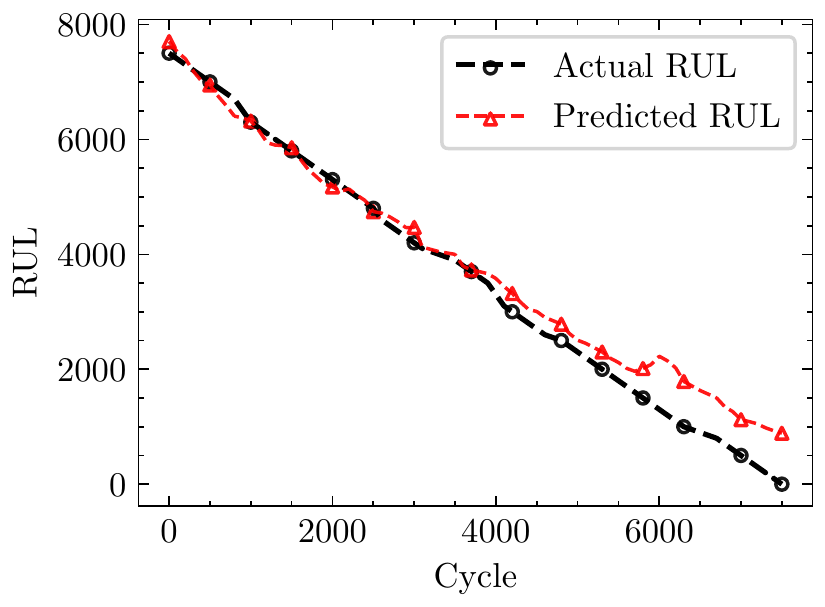} 
  } 
  \subfigure[Cell 8]{
  \label{fig:rul:cell8}
  \centering
  \includegraphics[width=0.35\columnwidth]{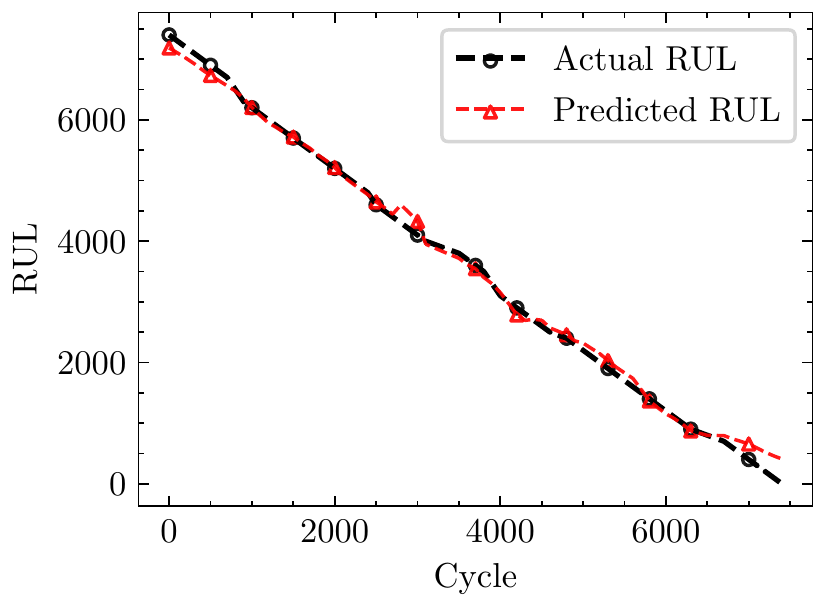} 
  } 
  \caption{The RUL estimation results of the eight cells in the Oxford dataset. The black circle dashed lines are actual RULs, and the red triangle dashed line predicted RULs with the DIICAN method.}
  \label{fig:rul_ox}
\end{figure}

\begin{table}[htbp]
  \caption{The RUL estimation of the eight cells in the Oxford dataset}
  \centering
    \setlength{\tabcolsep}{2.5mm}{
    \begin{tabular}{lcccc}
      \hline
      \textbf{Test Cell} & \textbf{MAE} & \textbf{MAPE} & \textbf{RMSE} & \textbf{$R^2$} \\
      \hline
      Cell 1 &152.4  & 6.2 & 191.7 & 0.993\\ 
      Cell 2 &280.9  & 4.5 & 363.4 & 0.973\\ 
      Cell 3 &108.0  & 8.0 & 149.5 & 0.996\\ 
      Cell 4 &249.9  & 13.3 & 304.9 & 0.945\\ 
      Cell 5 &560.3  & 16.0 & 582.9 & 0.817\\ 
      Cell 6 &210.3  &13.0 & 266.1 &0.956\\ 
      Cell 7 &294.9  & 12.9 & 400.5 & 0.967\\ 
      Cell 8 &104.0  & 5.7 & 138.6 & 0.996\\ 
      \hline 
      \end{tabular}} 
      \label{tab:rul_ox}
\end{table} 

\subsection{Training of DIICAN}
The DIICAN is trained and validated with the Pytorch module. We set the loss function as the mean absolute error (MAE). In order to achieve a rapid convergence and avoid overfitting, AdamW optimizer \cite[]{loshchilov2017decoupled} is used to train the unified method. The dropout technology \cite[]{srivastava2014dropout}, which decides the fraction of neurons to be excluded from training, is used as a regularization method to reduce the chances of overfitting. The initial learning rate, batch size and epochs are set to 0.0008, 64 and 100, respectively. The dropout rate is selected as 0.4. The hyper-parameters of the DIICAN method are presented in Table~\ref*{tab:para}.

And a leave-one-out validation scheme is used, whereby each cell is used once as a test set while the data from the remaining cells form the training set. In order to improve the learning process of the DIICAN method, all the battery feature data are scaled to the range of (0, 1) with the min-max normalization,
\begin{equation}
  \tilde{\mathbf{X}} = \frac{\mathbf{X}-min(\mathbf{X})}{max(x)-min(x)}
\end{equation}
where $max(X)$ and $min(X)$ are the maximum and minimum values of the variable $X$. 

\section{Results and performance analysis}\label{sec:results}
\subsection{Performance evaluation metrics}
To quantitatively evaluate estimation performances of the proposed method, four different evaluation metrics are selected: Mean Absolute Error (MAE), Mean Absolute Percentage Error (MAPE), Root Mean Squared Error (RMSE) and coefficient of determination ($R^2$) score. These metrics are described mathematically below:
\begin{align}
M\!AE &= \frac{1}{n} \sum_{i=1}^n \left|y_i-\tilde y_i\right| \\
M\!APE &= \frac{1}{n} \sum_{i=1}^n \frac{\left|y_i-\tilde y_i\right|}{y_i}\\
RM\!SE &= \sqrt{\frac{1}{n} \sum_{i=1}^n (y_i-\tilde y_i)^2}  \\
R^2 &= 1 - \frac{\sum_{i=1}^n (y_i - \tilde y_i)^2}{\sum_{i=1}^n (\tilde y_i - \overline y)^2}
\end{align}
where $n$ is the number of data points, $\tilde y_i$ and $y_i$ denote the estimated value and the real value respectively, and $\overline y$ is the mean value of observed $y_i$. RMSE evaluates performance by assigning more weights to large estimation errors, whereas MAE and MAPE consider all the estimation errors with same weights.

\subsection{Inter-cycle SOH-RUL estimation}
The real SOH and the predicted SOH are shown in Fig.~\ref{fig:soh_ox}. According to Fig.~\ref{fig:soh_ox} we can easily observe that the proposed DIICAN method is capable of forecasting battery SOH evolution. To further clarify the estimation performance, the performance metrics are presented in Table~\ref{tab:soh_ox}. 

The RUL estimation results of the eight cells are shown in Fig.~\ref{fig:rul_ox}. The comparison shows that the predicted RUL curve has a good agreement with the real RUL curve. In addition, the performance metrics of the RUL estimation are listed in Table~\ref{tab:rul_ox} for eight cells. And it should be noted that the total number of cycles for different batteries varies greatly. Thus, the Cell 4, Cell 5 and Cell 6 have the worse forecasting performance due to the shorter cycle life in regards to the other five batteries. 

Therefore, based on the above performance analysis, the proposed DIICAN method can accurately track the pattern of battery state evolution without a priori knowledge and can archive good accuracy for the battery health diagnosis.
\begin{figure}[h]
  \flushleft
  \subfigure[MAE]{
  \centering
  \includegraphics[width=0.3\columnwidth]{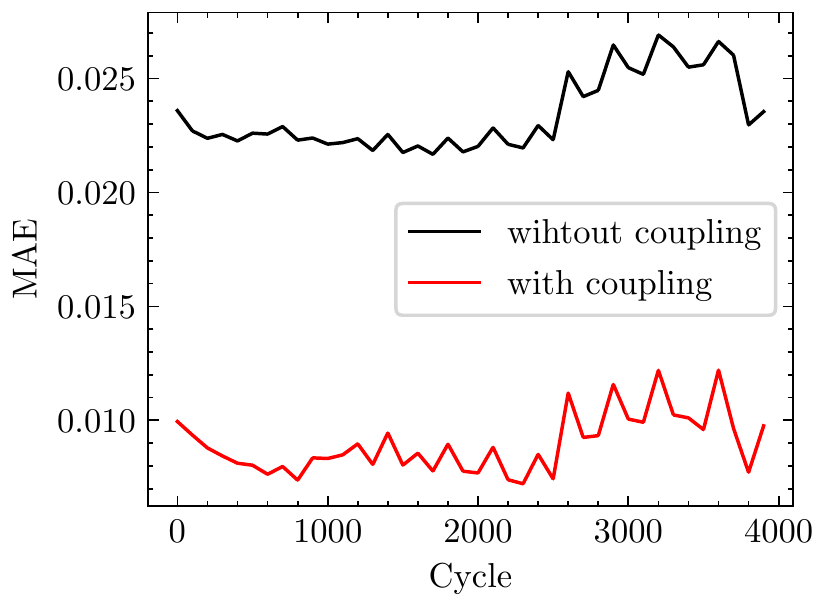} 
  } 
  \subfigure[RMSE]{
  \centering
  \includegraphics[width=0.3\columnwidth]{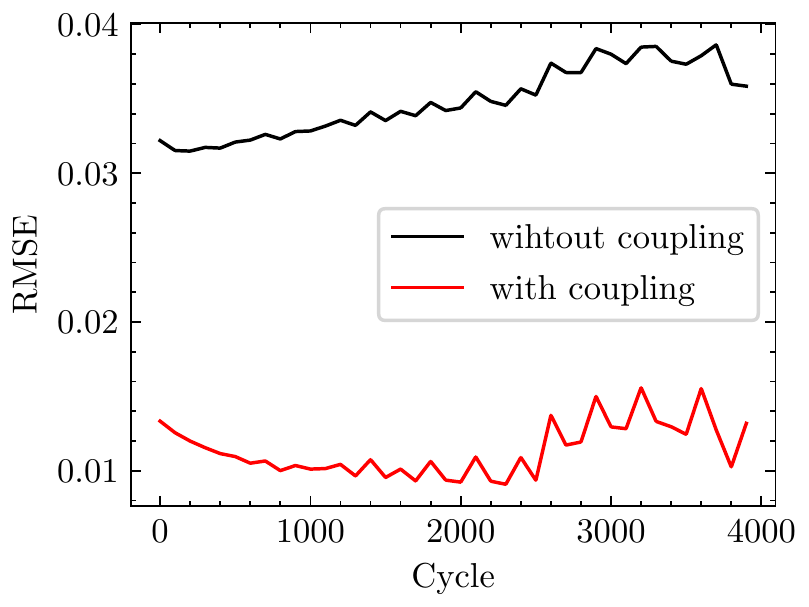} 
  } 
  \subfigure[$R^2$]{
  \centering
  \includegraphics[width=0.3\columnwidth]{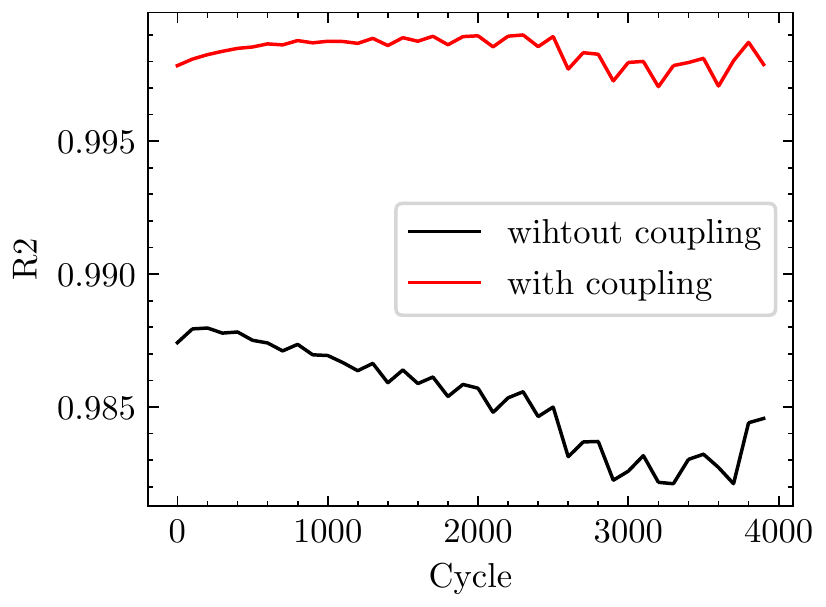} 
  } 
  \caption{The comparison of the SOC estimation results over the whole lifespan with state-coupling and without state-coupling. Three performance metrics, MAE, RMSE, and $R^2$ are presented.}
  \label{fig:soc_statistical}
\end{figure}

\subsection{Intra-cycle state-coupled SOC estimation}
Since battery aging can strongly affect its external properties, the SOC estimation demands the accurate SOH estimation.  The proposed DIICAN incorporates the inter-cycle hidden states to account for the influence of battery aging on the SOC estimation.
\begin{figure}[h]
  \centering
  \subfigure[SOH=0.95, RUL=4400]{
  \label{fig:soc:rul4400}
  \centering
  \includegraphics[width=0.4\columnwidth]{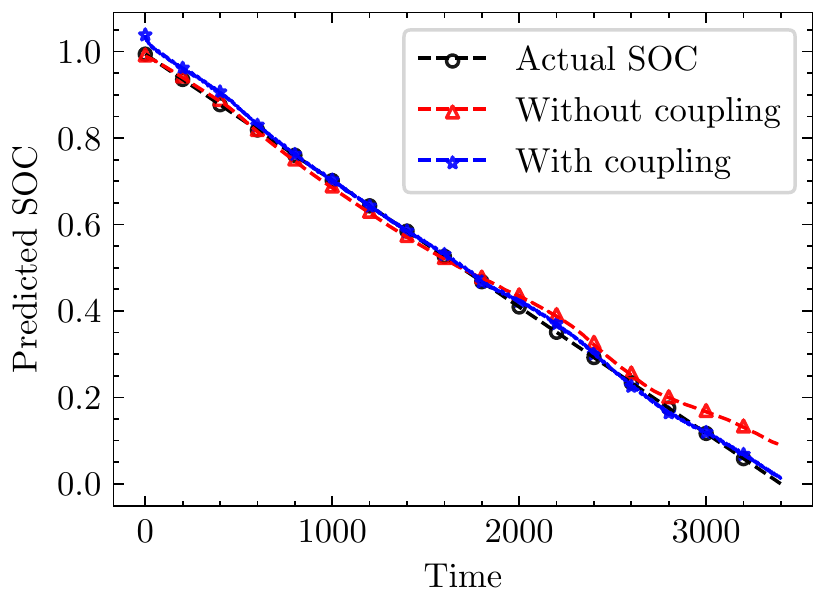} 
  } 
  \subfigure[SOH=0.93, RUL=3800]{
    \label{fig:soc:rul3800}
  \centering
  \includegraphics[width=0.4\columnwidth]{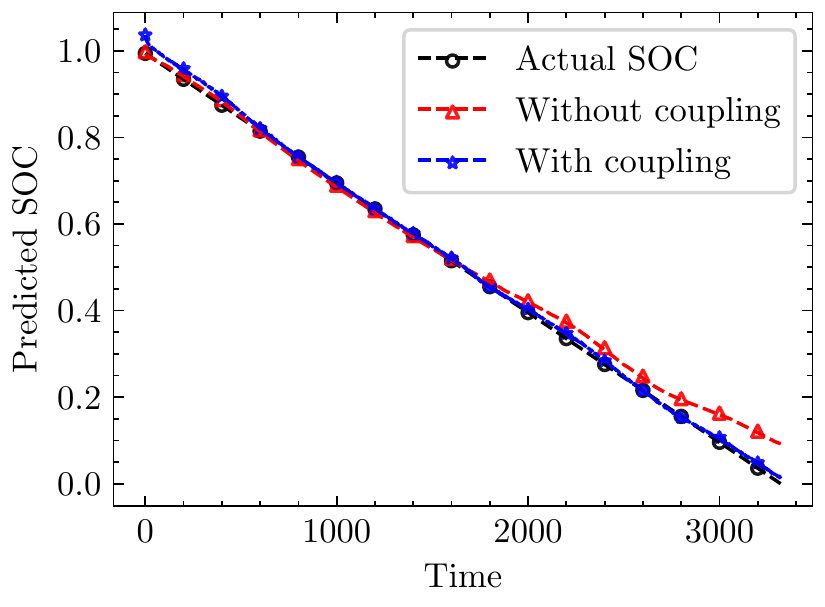} 
  } 

  \subfigure[SOH=0.90, RUL=3000]{
    \label{fig:soc:rul3000}
  \centering
  \includegraphics[width=0.4\columnwidth]{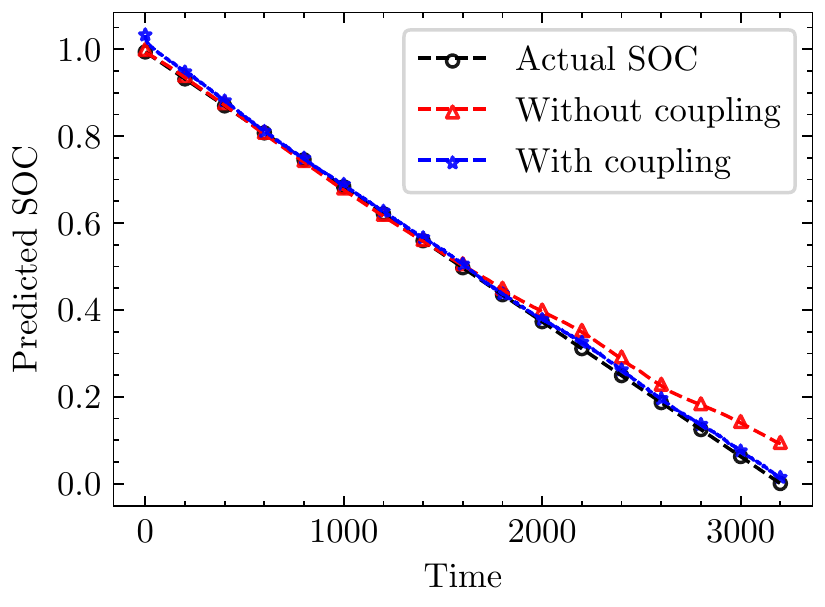} 
  } 
  \subfigure[SOH=0.87, RUL=2400]{
    \label{fig:soc:rul2400}
  \centering
  \includegraphics[width=0.4\columnwidth]{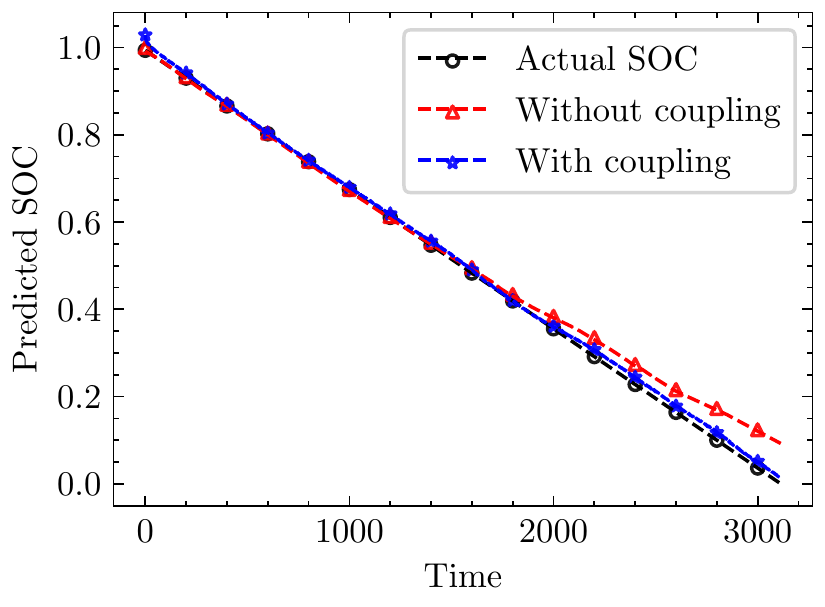} 
  } 

  \subfigure[SOH=0.83, RUL=1100]{
    \label{fig:soc:rul1100}
  \centering
  \includegraphics[width=0.4\columnwidth]{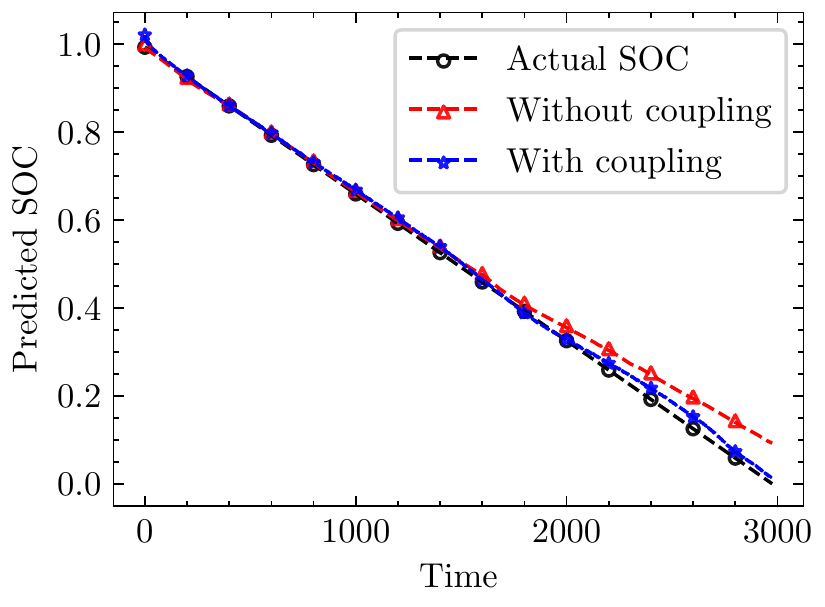} 
  } 
  \subfigure[SOH=0.81, RUL=500]{
    \label{fig:soc:rul500}
  \centering
  \includegraphics[width=0.4\columnwidth]{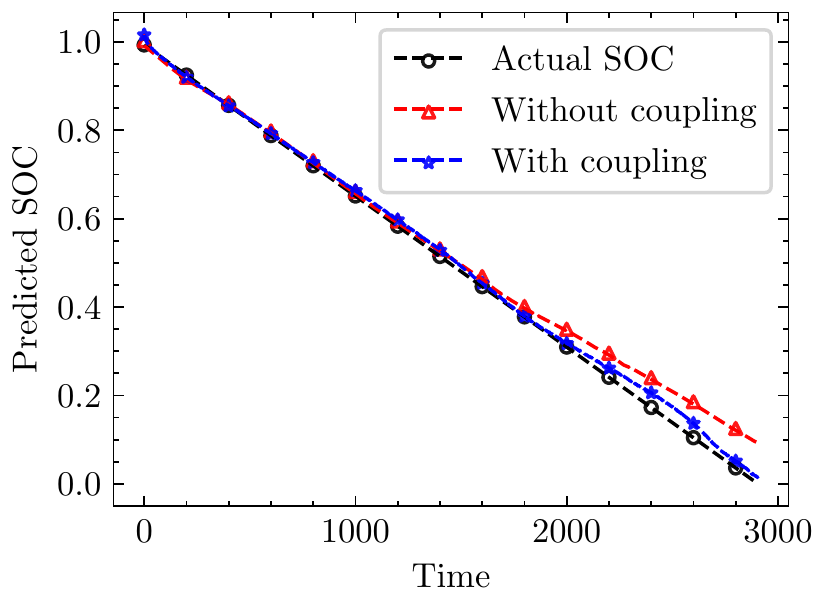} 
  } 
  \caption{The SOC estimation results. The black circle dashed lines are actual SOCs, the red triangle dashed line are estimated SOCs without state-coupling and the blue pentagram dashed line are estimated SOCs with state-coupling.}
  \label{fig:soc}
\end{figure} 

To verify its effectiveness, the SOC estimation results are shown in Fig.~\ref{fig:soc_statistical}.  As illustrated, it can be clearly seen that the SOC estimation accuracy clearly increased after capacity calibration. The MAE decreased from 2.34\% to 0.939\%, the RMSE decreased from 3.48\% to 1.12\% and $R_2$ score increased from 0.985 to 0.998 over the whole lifespan. In more detail, the SOC estimation results at 95\%, 93\%, 90\%, 87\%, 83\% and 81\% SOHs are shown in Fig.~\ref{fig:soc}, respectively. 

Neglecting the battery aging inevitably results in the inaccurate SOC estimation and the SOC estimation steadily drifts away without the discharge capacity calibration. It is clear that the capacity calibration can diminish the error of the SOC estimation caused by the battery degradation. Therefore, the SOC estimation performance is well preserved over the battery lifespan.

\subsection{Degradation pattern and attention weight}
Attention mechanisms can improve the model interpretability. As illustrated in Eq.~\ref{eqa:att_normal}, the SDA takes embedded vectors $\mathbf{S}_{0}$ of the initial cycle, temporal representations $\mathbf{h}_{i}$ of historical cycle feature and $\mathbf{h}_{i}^1\ominus \mathbf{S}_{0}$ to compute batteries' state degradation degree. 
\begin{figure}[h]
  \centering
  \includegraphics[width=0.35\columnwidth]{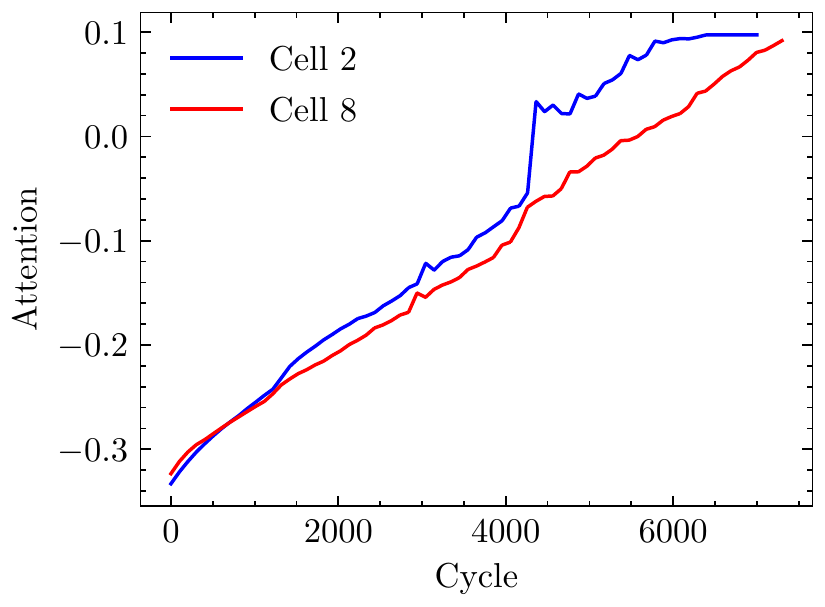}
  \caption{The attention weights of Cell 2 and Cell 8.}
  \label{fig:atten}
\end{figure} 

Taking Cell 2 and Cell 8 as examples to verify the effectiveness of the attention mechanism, attention weights are extracted from the forward propagation process of the trained model to present the intrinsic characteristics of attention mechanisms. The capacity degradation trajectories of those two cells are shown in Figs.~\ref{fig:soh:cell2} and ~\ref{fig:soh:cell8}. The capacity of Cell 8 degrades slowly and smoothly, while Cell 2 has bumpy trajectories with a sudden reduction in capacity at the 4300-th cycle.  In Fig.~\ref{fig:atten}, it can be seen that with the increase of cycle life, the attention weight increased gradually, which reflects the degree of battery aging. In addition, the attention weight of Cell 2 catches the sudden reduction of capacity. Overall, the results demonstrate that the proposed method can extract the pattern battery state degrades accurately and account for battery degradation degree sensitively.

\section{Conclusion}
\label{sec:conclusion}
Accurate SOC, SOH and RUL estimation is of great significance in BMS. In this paper, we have developed a dual time-scale state-coupled co-estimation method to estimate the SOC, SOH, and RUL. The inter-cycle SOH-RUL Estimation is applied to characterize battery degradation. Relying on the temporally structured recurrent module, the proposed method can extract the pattern how battery state evolves so that it is able to realize SOH and RUL estimation over the whole lifespan. Meanwhile, the effectiveness of the state degradation attention unit has been verified, which has an acute traceability to batteries' state degradation degree. And to compensate for the influence of battery aging, the inter-cycle hidden states which embodies battery degradation-related state is incorporated in the SOC estimation. With state coupling, the estimated SOC could track the reference over the whole SOC region well. In summary, the proposed method can yield sustained performance and realize accurate SOC, SOH and RUL co-estimation throughout the lifespan.

The battery state co-estimation algorithm is very useful for the intelligent modern BMS. The current research studies are mostly focused on battery state co-estimation for battery cells. The scalability of the battery state co-estimation algorithm from cell to module or even pack level demands more attention. In addition, when batteries are grouped, clustered or piled up, due to the inconsistency of battery cells, effectively achieving cell balance with the battery state co-estimation is critically demanding research work for the development of intelligent BMS.

\section*{Acknowledgments}
Xin Chen acknowledges the funding support from the National Natural Science Foundation of China under grant No. 21773182 and the support of HPC Platform, Xi’an Jiaotong University.

\printcredits

\bibliographystyle{unsrtnat}
\bibliography{mybibfile_1}

\bio{}
\endbio

\endbio

\end{document}